# Toward One-Second Latency: Evolution of Live Media Streaming

Abdelhak Bentaleb, *Member, IEEE,* May Lim, Mehmet N. Akcay, Ali C. Begen, *Senior Member, IEEE,* Sarra Hammoudi, and Roger Zimmermann, *Senior Member, IEEE*

*Abstract*—This survey presents the evolution of live media streaming and the technological developments behind today's IP-based low-latency live streaming systems. Live streaming primarily involves capturing, encoding, packaging and delivering real-time events such as live sports, live news, personal broadcasts and surveillance videos. Live streaming also involves concurrent streaming of linear TV programming off the satellite, cable, over-the-air or IPTV broadcast, where the programming is not necessarily a real-time event.

The survey starts with a discussion on the latency and latency continuum in streaming applications. Then, it lays out the existing live streaming workflows and protocols, followed by an in-depth analysis of the latency sources in these workflows and protocols. The survey continues with the technology enablers, low-latency extensions for the popular HTTP adaptive streaming methods and enhancements for robust low-latency playback. An entire section is dedicated to the detailed summary and findings of Twitch's grand challenge on low-latency live streaming. The survey concludes with a discussion of ongoing research problems in this space. We expect this survey to be the one-stop reference for those who would like to learn how low-latency live streaming has evolved and works today, and what further developments could happen in the future.

## I. INTRODUCTION

Video services, including video streaming, constitute more than 65% of the Internet's traffic in 2022 – a 24% increase over 2021 [180]. One major contributing factor is the significant growth in *live* video streaming, which grew from constituting less than 1% of the Internet's traffic in 2015 to almost 18% in 2022 [188]. These trends are powered by the proliferation of mobile devices, deployment of faster cellular and wireless networks, and advancements in audio-video capturing, editing and compression schemes. Moreover, the COVID-19 pandemic had temporarily put many cities around the world into lockdown, forcing millions of people to stay at home. This has contributed to an increase in live streaming by 10% in 2020 alone for well-known online over-the-top (OTT) streaming platforms [62].

A. Bentaleb is with Concordia University, Gina Cody School of Engineering and Computer Science, Canada (e-mail: abdelhak.bentaleb@concordia.ca).

M. Lim and R. Zimmermann are with the School of Computing, National University of Singapore, Singapore (e-mail: {maylim,rogerz}@comp.nus.edu.sg).

M. N. Akcay and A. C. Begen are with Ozyegin University, Istanbul, Turkey (e-mail: necmettin.akcay@ozu.edu.tr; ali.begen@ozyegin.edu.tr).

S. Hammoudi is with University of Bordj Bou Arreridj (e-mail: sarra.hammoudi@univ-bba.dz).

### A. What Constitutes Live Streaming

Let us start with the general definition of streaming. Streaming is the continuous transmission of media such as video, audio and metadata from a server to a client and its simultaneous consumption by the client. In this definition, the critical term is "simultaneous" as we would not call it streaming if one downloaded a media file and played it after the download was completed (this would be instead called download-and-play). There are two implications here. First, the transmission rate must (loosely or tightly) match the consumption rate in order to provide uninterrupted playback. That is, the client must not run out of data (a situation referred to as buffer underrun), where the playback would rebuffer, or accept more data than it can hold in its buffer (a situation referred to as buffer overrun), where any excess data would be discarded. Second, the client's consumption rate is limited by not only the bandwidth availability but also the real-time constraints. That is, the client cannot fetch or the server cannot push the media that is not available (*i.e.*, captured, encoded and packaged) yet.

In this survey, live streaming refers to streaming not only live content but also linear and personal broadcasts, and surveillance videos. Live content includes any event or program happening in real time, such as live sports and live news programming. Linear broadcasts include programs on a TV channel that consist of live (*e.g.*, live sports) as well as pre-recorded (*e.g.*, movies and TV shows) content. In the case of linear broadcast, liveness in live streaming refers to the real-time streaming of the broadcast content — not necessarily to the nature of the content. On the other hand, personal broadcasts include live captured content from personal devices such as mobile phones and gaming consoles, and surveillance videos include live video captured by public safety, security and traffic cameras.

### B. Latency and Latency Continuum

A viewer watching a football game, or breaking news, would be an example of live streaming consumption. One critical constraint in live streaming is the end-to-end (E2E) *latency* (also termed glass-to-glass latency), which is defined as the time gap between the capture of media (from a camera lens or microphone) and its playback (through a display or speaker).

How much latency is acceptable depends on the use case and application requirements. Fig. 1 depicts the continuum of different E2E latency values that are achievable or required by different applications, platforms and streaming configurations.

Based on today's common understanding within the industry, we divide OTT live streaming into five broad categories based on the target E2E latency, noting that this categorization is subject to change based on future practices:

(a) *High Latency*: E2E latency is 45 or more seconds. The common application of this category is the one-way streaming of live events to large audiences where latency is relatively less important than the scalability and robustness of the service offering.
(b) *Typical Latency*: E2E latency ranges between 10 and 45 seconds. This latency range is commonly achieved by the majority of OTT services using HTTP adaptive streaming (HAS). It is also used as a fallback state when some components of the E2E media delivery pipeline do not have the necessary support for low-latency live streaming.
(c) *Low Latency*: E2E latency ranges between one and 10 seconds. This level of latency is essential to have a user experience similar to traditional broadcast TV. Example event types for this category include premium live sports, financial news and eSports. Another example application is second-screen experiences, where an event can simultaneously be consumed on different devices.
(d) *Ultra-Low Latency*: E2E latency is a second or less. Interactive experiences such as live commentary, betting, in-game wagering, online gambling and auctioning are the main drivers in this category. Ultra-low-latency live streaming introduces new challenges because this level of latency is on the same scale as commonly observed network latency variations, often due to bufferbloat [158] and other network anomalies such as packet reordering.
(e) *Near-Real-Time Latency*: E2E latency is in the order of tens of, up to 100, milliseconds. This level of latency is essential for networked applications that incorporate live streaming but also require specific interactivity or control-feedback features. Videoconferencing, cloud gaming, remote control of hardware platforms like drones, vehicles or surgical robots and metaverse are example applications.

We will survey the technologies that enable the latency ranges (a)–(c) above. However, our primary focus will be the low-latency live (LLL) streaming applications, where the key goal is to keep the latency below 10, preferably five, seconds. In this survey, however, we will not go into the details of the protocols used to achieve ultra-low (d) and near-real-time (e) latencies. Instead, Section VI briefly touches on these topics and the preliminary discussions taking place in the new working group (called Media over QUIC or MOQ in short) in the Internet Engineering Task Force (IETF).

Today, an E2E typical latency of 10–45 seconds is practically achievable by the majority of live streaming services using HAS. However, such latency range is not low enough for certain content or use cases [57], [194]. To support low latency, extensions for the popular HAS methods, namely Dynamic Adaptive Streaming over HTTP (DASH) and HTTP Live Streaming (HLS), have been developed and they are known as Low-latency DASH (LL-DASH) [67] and Low-latency HLS (LL-HLS) [17]. Details of these HAS methods and their low-latency extensions are presented in Sections III-D1 and V.

Several recent technological advances have laid out the foundation for LLL streaming. Nevertheless, significant limitations and challenges still exist. Deploying an LLL streaming service requires a diligent examination and streamlining of every component in the media delivery pipeline. To this end, it is imperative to understand all the components and functionality that contribute to the E2E latency. Examining all the latency sources is one of the main goals of this survey.

### C. Experience Metrics in Live Streaming

Two important time-related metrics in live streaming are the *startup delay* and the *latency*, which are distinct from each other. The startup delay is the time lag from when the client joins a live stream (*i.e.*, the viewer presses PLAY) until it begins to play out the media, while the latency is the time lag from when a media frame is captured until the same frame is played out. As the name implies, the startup delay only impacts the beginning of a stream. However, the latency persists throughout the whole playback and may sometimes increase or decrease in parts of the playback. These metrics are also independent of each other. That is, the startup delay might be quite small or large, while the latency could be low or high.

The startup delay and latency are both influential metrics. Together with rebuffering events and the playback speed, they can significantly impact the viewer's quality of experience (QoE). The viewer QoE is a terminology that is introduced to describe a viewer's perception and it reflects how satisfied or dissatisfied the viewer is while watching a stream based on various factors. There are two essential techniques to evaluate the viewer's QoE: subjectively or objectively. A subjective QoE evaluation is performed through a user study, while an objective QoE evaluation is based on mathematical models such as ITU P.1203 (an open-source implementation is available in [177]) or the Yin model [230], which typically consider a weighted combination of the selected QoE factors. The eventual QoE value for a streaming session is generally a value in the same range as a mean opinion score (MOS, ITU P.800.1), which is between one ('bad') and five ('excellent').

Achieving a higher QoE often requires streaming higher-quality content with fewer rebuffering events. Higher-quality content means using a higher encoding bitrate due to higher resolution and/or frame rate, and a wider color gamut, *etc*. It may also mean a longer encoding time to allow the encoder to compress the media better using a longer look-ahead buffer or a multi-pass encoding approach. However, the time budget allocated for encoding is severely limited in the case of LLL streaming (as opposed to on-demand streaming). Thus, the encoders need to be well-designed for low-latency operation, and the number of bitrate-resolution pairs (also known as the bitrate ladder) offered to the streaming clients should be carefully selected. On the other hand, eliminating or at least reducing the occurrences of rebuffering events, which are common due to transient network conditions, is easier with more media being buffered in the playback buffer. However, there is a limit to this, too, as buffering more media increases the latency unless the streaming client employs an adaptive playback speed controller to keep the latency under control while reducing the risk of rebuffering. In Table I, we summarize the key QoE metrics for LLL streaming, and in Section VIII, we demonstrate how all these different experience metrics can be combined to compute a single QoE value.

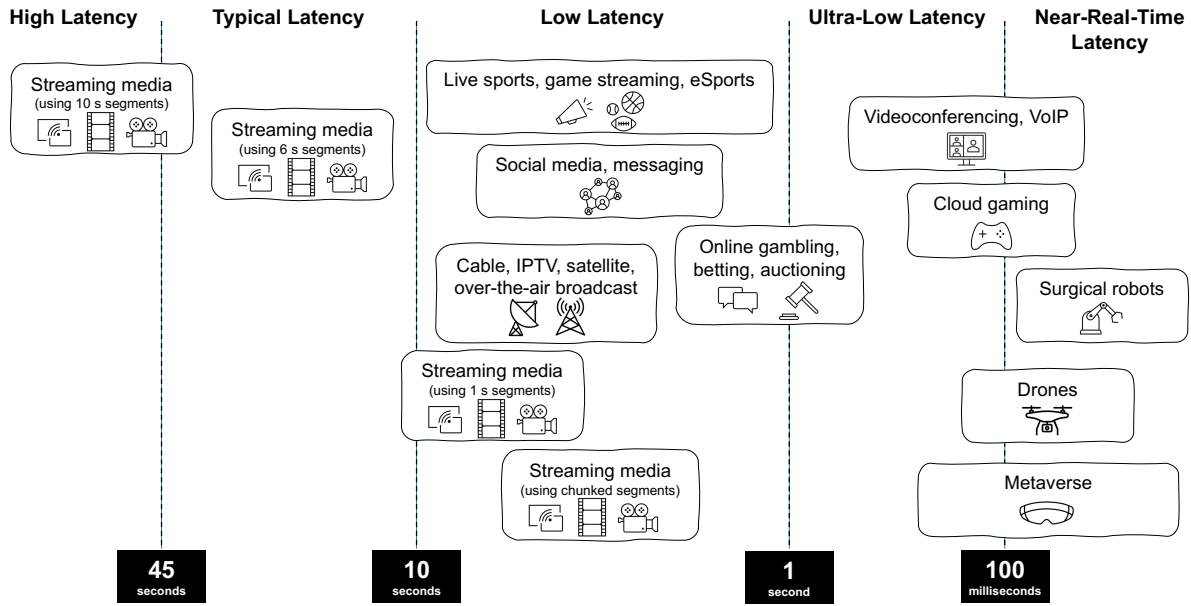

Fig. 1: Continuum of different E2E latency values. Latency values are not to scale.

Specifically in wireless networks, technologies related to Ultra Reliable Low Latency Communications (URLLC), such as finite blocklength coding, short transmission time interval, non-orthogonal multiple access and mini-slots [7], are instrumental for achieving low latency. URLLC, a critical component in 5G and emerging 6G networks, aims to provide high reliability and low latency for mission-critical applications. However, these technologies and related metrics fall outside the scope of this survey. Instead, we focus on the most widely used metrics in the literature (refer to Table II), specifically emphasizing QoE-related metrics such as E2E latency, startup delay, quality, rebuffering and playback speed, to ensure a concise discussion.

TABLE I: Definitions of key QoE metrics in the context of LLL streaming.

| Metric | Definition |
|---|---|
| E2E Latency | Time gap between media capture and playback (s). |
| Startup Delay | Time from the client joins the stream until playback starts (s). |
| Quality | Codec, encoding bitrate (Kbps or Mbps) and other parameters. |
| Rebuffering | Event when the playback is paused due to buffer depletion (s). |
| Playback Speed | Actual playback speed, compared to the normal speed (1x). |

## II. CONTRIBUTIONS, SCOPE AND STRUCTURE OF THE SURVEY

### A. Contributions

In the rapidly evolving landscape of LLL streaming, achieving low latency remains a critical challenge that directly impacts user experience and engagement. While some surveys (highlighted in Table II) have explored various facets of live streaming technologies and protocols, there remains a significant gap in comprehensively addressing the E2E latency sources and the latency mitigation strategies. This survey bridges this gap by providing an in-depth analysis of latency across the entire media delivery pipeline—from content preparation and delivery to consumption. Our survey distinguishes itself by providing an exhaustive analysis of E2E latency sources across the entire live streaming workflow. We systematically categorize and examine the components contributing to latency, including media content preparation, delivery and consumption. Furthermore, we explore both the established and emerging protocols, offering a critical comparison of their effectiveness in reducing latency. By integrating practical case studies, such as Twitch's grand challenge on low-latency live streaming, and discussing future directions like Media over QUIC (MOQ), our survey serves as a one-stop reference for researchers and practitioners aiming to advance LLL streaming technologies.

Table II presents a detailed comparison between this survey and related surveys.

### B. Scope

The primary focus of this survey is (*i*) to provide a comprehensive anatomy of the components in live and low-latency live streaming systems, and how these components

TABLE II: Comparison to related surveys. Our focus is to provide an in-depth analysis of latency across the entire media delivery pipeline—from content preparation and delivery to consumption.

| Survey Ref. | Focus Area | Scope |
|---|---|---|
| [6] (2024) | Energy consumption in video streaming | Analyzes the environmental impact of streaming technologies, emphasizing energy-efficient solutions and sustainability in video delivery systems. |
| [207] (2023) | Immersive video streaming | Covers the end-to-end pipeline for immersive video, including acquisition, compression, transmission and display, with a focus on immersive-specific challenges. |
| [119] (2022) | MEC and its role in video streaming | Examines multi-access mobile edge computing (MEC) frameworks for reducing latency and enhancing QoE in video streaming applications. |
| [65] (2022) | Computation-driven live streaming | Focuses on computational optimizations for live streaming, such as encoding, transcoding and resource allocation, to improve efficiency and reduce latency. |
| [129] (2021) | Cloud-based video streaming | Investigates the role of cloud computing in video storage, processing and delivery, with an emphasis on scalability and infrastructure. |
| [115] (2021) | Video streaming in vehicular networks | Reviews resource allocation strategies for video streaming in vehicular environments, addressing challenges specific to vehicle-to-everything (V2X) communication. |
| [228] (2020) | ABR for 360° video | Reviews ABR techniques tailored for 360° video streaming, addressing challenges in delivering immersive video experiences. |
| [78] (2019) | Acquisition, transmission and display of 360° video | Covers the complete pipeline for 360° video streaming, including capture, compression, transmission and rendering. |
| [40] (2019) | ABR for HAS | Reviews bitrate adaptation techniques for HAS, focusing on optimizing QoE through dynamic bitrate selection. |
| [124] (2017) | ABR for HAS | Surveys ABR algorithms and frameworks for HAS, focusing on bitrate adaptation in response to network conditions. |
| [181] (2015) | QoE for HAS | Discusses subjective and objective methods for assessing QoE in video streaming, with a focus on HAS. |

affect E2E latency, and (*ii*) to describe the state-of-the-art techniques and enhancements that have been proposed to date to achieve robust playback at low latencies. The survey is structured as follows:

In Section III, we first introduce the general architecture of today's live streaming systems and describe the different stages within this architecture, namely media contribution, ingest and distribution. Then, we list the protocol options for each stage and discuss their pros and cons. Specifically, we cover the push-based protocols for the contribution and pull-based and peer-to-peer (P2P) protocols for the distribution stages. Finally, we summarize the challenges for all these protocols.

In Section IV, we dissect the E2E latency into three serial parts for a more organized examination: media content preparation, delivery and consumption latency. We provide an in-depth analysis of the components contributing to the latency for each part. Then, we highlight improvements for each component to meet the LLL streaming requirements.

In Section V, we introduce two key technologies enabling LLL streaming: chunked encoding/packaging and chunked delivery. We use the Common Media Application Format (CMAF) standard and HTTP/1.1's chunked transfer encoding mode as examples. Then, we present the low-latency extensions for the existing HAS protocols, namely LL-DASH and LL-HLS, and show how the same chunked-encoded media can be efficiently served using these two extensions. The section then continues with a discussion on the High Efficiency Streaming Protocol (HESP) and the proposed Low-Latency Low-Delay Extensions for DASH (L3D-DASH).

In Section VI, we present a brief overview of the developments in using real-time communication protocols together with DASH to improve interactivity in LLL streaming. We also summarize the preliminary discussions that led to the forming of the new Media over QUIC (MOQ) working group in the IETF.

In Section VII, we discuss the main components of a media player and discuss bandwidth measurement, rate adaptation (adaptive bitrate, ABR), playback speed control and (playback) buffer-management algorithms designed for low latency. The section also provides an evaluation of the implementations for existing open-source media players.

In Section VIII, we present the summary and findings of Twitch's grand challenge on low-latency live streaming [206]. This summary illustrates the best practices to prototype an E2E LLL streaming solution and test the individual components in the delivery workflow.

In Section IX, we present the lessons learned that summarize the key takeaways from this survey.

Finally, Section X concludes the survey by summarizing the lessons learned and future directions for LLL streaming.

The structure of the survey is highlighted in Fig. 2. The list of the abbreviations used frequently in this survey is also given in Table III.

*C. Pointers for Out-of-Scope Subjects*

While we cover many topics on low-latency live streaming in this survey, the following subject areas (for which we provide some references for interested readers) are out-of-scope:

- Two-way, interactive videoconferencing applications: [52], [112], [141], [175].
- Cloud gaming [4], [48], [72], [87].
- QoE management, measurement and monitoring: [9], [23], [98], [181], [210].
- Stream manipulation, content steering and ad insertion: [18], [19], [60], [76], [185]–[187].

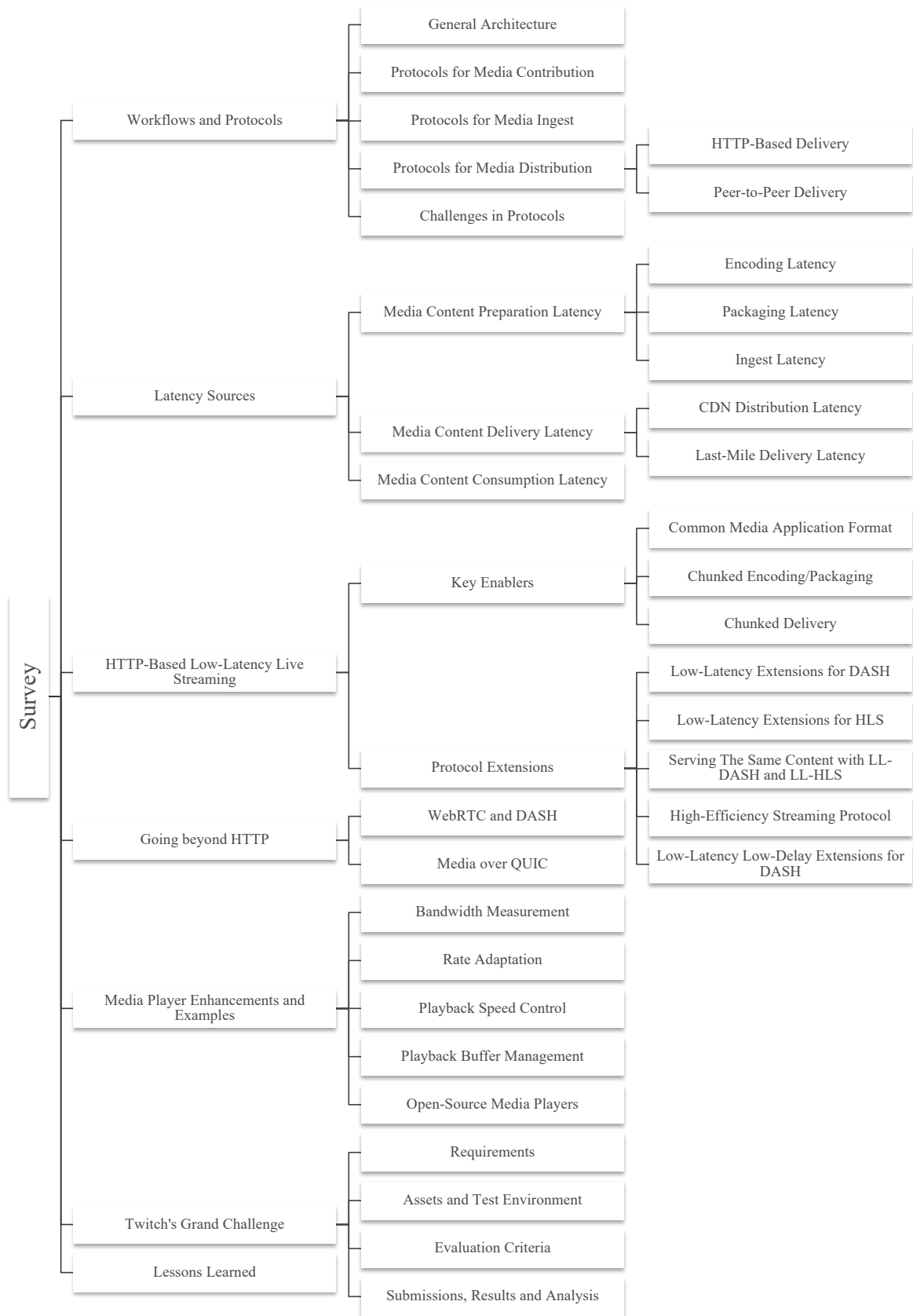

Fig. 2: Structure of the survey.

TABLE III: List of the frequently used acronyms and abbreviations.

| Description | Acronym |
| --- | --- |
| Adaptive bitrate | ABR |
| Advanced Video Coding (H.264) | AVC |
| Chunked transfer encoding | CTE |
| Common Media Application Format | CMAF |
| Common Media Client Data | CMCD |
| Common Media Server Data | CMSD |
| Content delivery network | CDN |
| DASH Industry Forum | DASH-IF |
| Digital rights management | DRM |
| Dynamic Adaptive Streaming over HTTP | DASH |
| Encrypted Media Extensions | EME |
| End-to-end | E2E |
| Group of pictures | GoP |
| High Efficiency Streaming Protocol | HESP |
| HTTP adaptive streaming | HAS |
| HTTP Live Streaming | HLS |
| ISO Base Media File Format | ISO BMFF |
| Low-latency low-delay DASH | L3D-DASH |
| Low-latency DASH | LL-DASH |
| Low-latency HLS (community version) | LHLS |
| Low-latency HLS (official Apple version) | LL-HLS |
| Low-latency live | LLL |
| Media content consumption latency | MCCL |
| Media content delivery latency | MCDL |
| Media content preparation latency | MCPL |
| Media over QUIC | MOQ |
| Media presentation description | MPD |
| Media Source Extensions | MSE |
| Moving Picture Experts Group | MPEG |
| Microsoft Smooth Streaming | MSS |
| Network Time Protocol | NTP |
| Over-the-top | OTT |
| Peer-to-peer | P2P |
| Per-title encoding | PTE |
| Point of presence | PoP |
| Quality of experience | QoE |
| Quality of service | QoS |
| Real-time Messaging Protocol | RTMP |
| Real-time Transport Protocol | RTP |
| Real-time Streaming Protocol | RTSP |
| Round-trip time | RTT |
| RTP Control Protocol | RTCP |
| Secure Reliable Transport | SRT |
| Transmission Control Protocol | TCP |
| User Datagram Protocol | UDP |
| World Wide Web Consortium | W3C |

## III. WORKFLOWS AND PROTOCOLS FOR LIVE STREAMING

In this section, we first introduce the general architecture of today's live streaming systems and describe the different stages within this architecture, namely media contribution, ingest and distribution. Then, we list the protocol options for each stage and discuss their pros and cons. Specifically, we cover the push-based protocols for the contribution and pull-based and P2P protocols for the distribution stages. Finally, we summarize the challenges for all these protocols.

### A. General Architecture

A typical E2E live streaming architecture is shown in Fig. 3, consisting of three main stages: *media contribution*, *media ingest* and *media distribution*. The contribution stage involves content acquisition and production. As the content is acquired from a source (*e.g.*, via cameras at a studio or stadium), it is first encoded by a contribution encoder to produce very high-quality output at a high bitrate. The output of the contribution encoder is then input to a multi-bitrate distribution encoder (also called an OTT encoder or transcoder). For the transmission from the contribution encoder to the distribution encoder, one can use a specialized media transport protocol (see Section III-B for some examples). One can also use plain multicast if there are multiple distribution encoders running in parallel for load balancing or redundancy purposes. If the content is acquired from an artificial source (*i.e.*, , it is AI-generated content (AIGC) [49], [223], [224], as illustrated in Fig. 3), there may be an additional delay due to the real-time synthesizing process before the AIGC is prepared for distribution. However, this delay is orthogonal to the latency aspects we study in this survey.

The second (ingest) stage involves content packaging, encryption and transmission to an origin server. Out of the incoming high-quality contribution feed, the distribution encoder generates multiple representations (at different bitrates and resolutions using one or more codecs) according to a preset bitrate ladder. It then packages them into one or more delivery formats, such as DASH and HLS, and optionally applies encryption. The packaged content that includes all the DASH/HLS media segments and the associated manifest files is then ingested to an origin server. There are many options for the ingest protocol, some of which are described in Section III-C.

Finally, the distribution stage involves content delivery and consumption. The packaged content is distributed from the origin server to media players connected through different last-mile access networks. When using HTTP for distribution, one or more content delivery networks (CDN) can be used to shield the origin server and scale the distribution to millions of media players around the globe. As is the case with the other stages, there are many options one can use for distributing the content. Some examples are provided in Section III-D.

The choice and configuration of the various components, such as the contribution and distribution encoders, packagers, contribution, ingest and distribution protocols, as well as rate-adaptation, buffer-management and other playback-related algorithms running on the media player, all affect the E2E latency and robustness of the playback. In the remainder of this section, we summarize various protocol options for media contribution, ingest and distribution, along with their advantages and disadvantages.

### B. Protocols for Media Contribution

Protocols used for media contribution are generally push-based protocols where the server/encoder *pushes* the media. A few traditional protocols were designed in the early days of streaming over the Internet and have been in use for many years. Nonetheless, the last decade has seen an emergence of new standards-based and proprietary protocols. These were designed for reliable, low-latency live media transmission over best-effort (*i.e.*, unmanaged) networks. The end goal for these protocols has been to reduce operational costs and increase deployability. We summarize these protocols below.

- **Real-Time Messaging Protocol (RTMP)** (RFC 7425) was created by Adobe as a delivery protocol for streaming

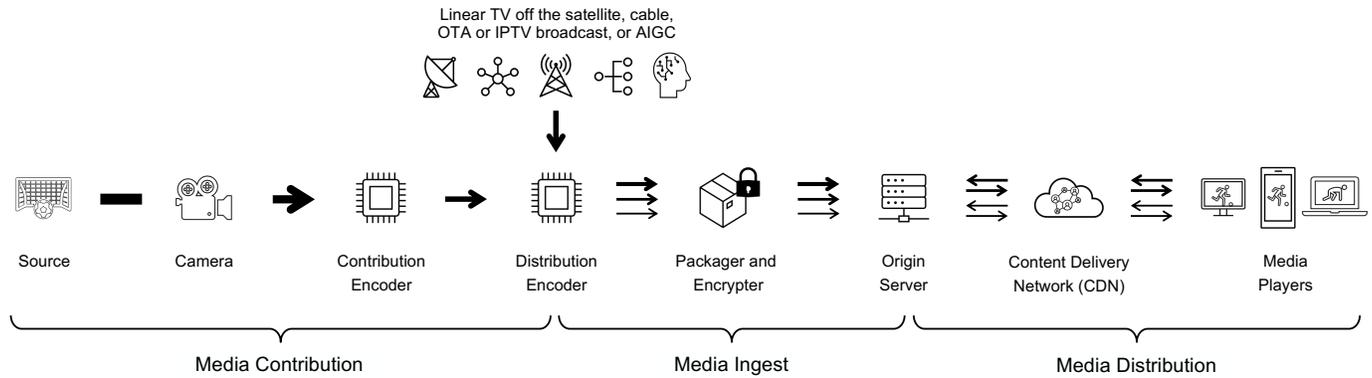

Fig. 3: A typical E2E live streaming architecture. The thickness of the arrows denotes the bitrate for that particular media stream (not to scale). There are usually multiple media streams (audio, video, metadata, subtitles, *etc.*) acquired from the source (not shown).

video and audio over the Internet. It uses a persistent TCP connection to transmit media data from a source to a single receiver with Adobe Flash Player support. It was initially designed for both video contribution and distribution. However, because Flash was (*i*) a proprietary technology, (*ii*) increasingly vulnerable to cyber attacks, and (*iii*) deprecated and not allowed to work in modern browsers, playing media over RTMP is no longer supported on many end-points. While Adobe announced the end-of-life for its Flash Player back in 2017 [5], RTMP remains popular and many content platforms continue to use it for media contribution.

- **Real-Time Transport Protocol (RTP)** (RFC 3550) is a protocol developed for the real-time transmission of multimedia data in unicast or multicast mode. Although it is not required, RTP typically runs over UDP. Through its header, RTP supports payload type identification, sequence numbering and timestamping. RFC 3550 also defines the RTP Control Protocol (RTCP), whose primary purpose is to provide applications with minimal control and identification functionality. Additionally, RTCP provides a scalable monitoring service for RTP transport by using sender, receiver and extended reports (RFC 3611). While RTP is widely used as a standalone protocol in many multimedia applications, for media contribution, it is used as part of another protocol or standard, such as described below.
- **SMPTE ST 2022** [189] is a suite of standards developed by the Society of Motion Picture and Television Engineers (SMPTE) that focuses on professional quality video delivered over IP. The suite currently consists of eight standards. The ST 2022-2, 2022-3 and 2022-4 standards describe the carriage of MPEG-2 Transport Streams over IP. The ST 2022-1 and 2022-5 standards define a forward error correction technique for the carriage of real-time compressed and uncompressed (termed as High Bit Rate (HBR), 270 Mbps or higher) video/audio over IP, respectively. The ST 2022-6 standard defines a unidirectional IP-based protocol for the transport of real-time uncompressed video, audio and ancillary data typically found on serial digital interfaces (SDI). The ST 2022-7 standard is for protecting a video stream against data loss by seamlessly switching between multiple redundant streams of RTP packets. Finally, the ST 2022-8 standard specifies the use of and constraints for ST 2022-6 streams in conjunction with the timing model of ST 2110-10.
- **Secure Reliable Transport (SRT)** [96], [219] is an open-source protocol first released in 2013 to provide end-to-end security, resiliency and dynamic end-point adjustment based on real-time network conditions while delivering high-quality video. SRT has been shown to have high robustness against packet loss, jitter, latency and variable bandwidth. The development of SRT is managed by the SRT Alliance, which currently has 400+ members [95]. SRT runs over UDP and is primarily used for contribution, although some projects are underway to use it for distribution. In that case, due to the lack of support and integration in browsers, an SRT gateway will be needed to convert the incoming SRT stream into a format the browsers can use.
- **Reliable Internet Stream Transport (RIST)** [211], [219] is a standard developed by the Video Services Forum (VSF). It is a reliable, low-latency video transmission solution intended to be used over unmanaged networks. RIST uses RTP transport for the media packets where any missing packet is reported using RTCP feedback (RFC 4585) and then retransmitted using RTP retransmissions. RIST has multiple profiles defined so far (Simple, Main and Advanced) with different feature sets. RIST is also available in open source and has been implemented in many popular media software packages.
- **Zixi** [219], [237] is a proprietary congestion and network-aware protocol designed for media transmission over best-effort networks. It dynamically adjusts to varying bandwidth and includes a set of forward error correction (FEC) techniques to ensure reliable and low-latency delivery.

*C. Protocols for Media Ingest*

With some exceptions, requirements for the media ingest protocols are similar to those of the media contribution protocols. Therefore, most of the contribution protocols (RTMP, SRT, RIST, Zixi, *etc.*) can be used for media ingest with no or little changes. On the other hand, there are ingest protocols developed for HTTP-based workflows. There are also those proposed for personal live broadcasts (*i.e.*, there is not necessarily a media contribution stage). We summarize these protocols below.

- **Microsoft Smooth Streaming Live Ingest** is a legacy media ingest protocol developed by Microsoft as part of the Smooth Streaming product suite [148]. This protocol was widely used for both on-demand and live content for several years. With the advances in media codecs, distribution protocols and signaling of metadata and timed text, this protocol was later replaced by **Microsoft's Azure Media Services fragmented MP4 Live Ingest** protocol [147]. This new protocol supports service/encoder failover and redundancy to improve the robustness of the ingested media.
- **DASH-IF Live Media Ingest Protocol** [68] is a specification developed through the collaboration of many vendors and service providers. The specification defines two interfaces. Both interfaces use the HTTP POST (or PUT) method to transmit media objects from an ingest source to a receiving entity. The first interface describes the direct ingest of media formatted as a CMAF presentation to a receiving entity. The receiving entity, a packager or a cloud media service, can produce DASH and HLS segments along with the manifests with optional transcoding, stitching, (re)encrypting and watermarking. In the second interface, the media is already packaged in the DASH or HLS format. In this case, the media objects are transmitted to a receiving entity (such as an origin server or a cloud media service) along with the manifests. If desired, the receiving entity can do further manifest manipulation. Otherwise, it stores and serves the segments. Both interfaces support the carriage of video and audio data, timed metadata and timed text. The specification also presents the guidelines for synchronizing multiple ingest sources and discusses redundancy and failover behaviors. Some open-source tools are provided in [68].
- **Warp** [140] is a segmented live media transport protocol proposed by Twitch and is currently under discussion in the IETF. Warp takes any live input and maps it to individual QUIC streams based on the specific encoding. By adjusting the transmission priorities, Warp tries to deliver as much media as possible, given the latency requirements. Warp can be used for both media ingest and distribution. An open-source implementation and demo are available at [139].
- **RUSH** [173] is a bidirectional application-level protocol designed for live media ingest that runs on top of QUIC. RUSH's primary goal is to replace RTMP with support for new codecs, message types and multi-track support. Similar to Warp, RUSH also provides provisions for timely (but not necessarily fully reliable) media delivery. RUSH can also be used for both media ingest and distribution.
- **WebRTC-HTTP Ingestion Protocol (WHIP)** [155] is an HTTP-based protocol that allows Web Real-Time Communication (WebRTC)-based ingestion of media. A WHIP client producing and encoding live media sends an HTTP POST request to establish a session with the ingest end-point. Once the session is set up, media flows unidirectionally from the WHIP client to this ingest end-point, with no possibility of adding or removing media tracks. WHIP is currently an active working group document in the IETF and is close to publication. Several open-source implementations are also available (*e.g.*, [77], [143], [144]).

*D. Protocols for Media Distribution*

Media contribution and ingest protocols often run in point-to-point mode, although they can also run in point-to-multipoint mode for load balancing and redundancy purposes. The top priority for these protocols is reliability, *i.e.*, no-loss media transmission, since any data loss during media contribution or ingest introduces an artifact that will impact any viewer consuming this media content. In the case of live media, low-latency contribution and ingest also become important. On the other hand, media distribution protocols also run in mostly point-to-point mode. However, these protocols need to be able to scale from a few to millions of media players. While media distribution originates from an origin server, it is always shielded by other entities (caches, replication points, *etc.*) that distribute the content on behalf of the origin server. Next, we first present the details for using HTTP for media distribution (*i.e.*, HTTP adaptive streaming, HAS), which is by far the most popular approach today. Then, we briefly discuss peer-to-peer (P2P) delivery and give examples.

*1) HTTP-Based Delivery:* Before presenting how HAS works, let us briefly review HTTP's history, as the recent protocol changes are important to understand in the context of LLL streaming.

The Hypertext Transfer Protocol (HTTP) is an application-level protocol that has been in use by the World Wide Web (WWW) for over 30 years. The initial HTTP version (HTTP/0.9) was a simple protocol prototyped for raw data transfer from a server to a client. In 1996, RFC 1945 documented HTTP/1.0, where the initial protocol was improved by allowing messages to be in the format of MIME-like messages, containing metainformation about the data transferred and modifiers on the request/response semantics. RFC 1945 was published with some concerns since it did not fully specify what was mandatory or optional to implement. Moreover, it did not sufficiently consider the effects of hierarchical proxies, caching, the need for persistent connections and virtual hosts. As the Web has become quite popular, incomplete implementations caused interoperability issues, necessitating a protocol version change. Shortly after, in early 1997, RFC 2068 specified HTTP/1.1 with more stringent requirements to ensure reliable and interoperable implementations. Two years later, RFC 2068 was revised again to RFC 2616, which stayed in effect for a long time. In 2014, HTTP/1.1 was overhauled significantly and revised in RFCs 7230 through 7235. The latest revision to HTTP/1.1 was made in 2022 and the current standard is specified in RFCs 9110, 9111 and 9112.

The second major version of the HTTP, dubbed HTTP/2 (H2), was released in 2015 by the IETF in RFC 7540. H2 was derived mainly from the ideas inherited from Google's SPDY proposal, an experimental protocol first announced in 2009 [151] with the objective of reducing page load times. H2 includes many new features, such as stream multiplexing, server push, stream priority and stream termination. Instead of running multiple TCP connections in parallel for handling multiple simultaneous requests, as in the case of HTTP/1.1, H2 uses a single TCP connection to carry multiple request-response pairs (via multiple streams). Despite all the new features, H2 still uses TCP as the underlying reliable transport protocol and is subject to complications resulting from head-of-line (HoL) blocking. The current specification for H2 is RFC 9113.

The third major and latest HTTP version is HTTP/3 (H3), specified in RFC 9114. In contrast to HTTP/1.1 and H2 that use TCP, H3 is built on the QUIC protocol (RFC 9000), which provides a secure and reliable transport protocol over UDP. Like TCP, QUIC is a connection-oriented and stateful protocol that provides interaction between a client and server. QUIC allows an application to establish a connection quickly and uses flow-controlled streams multiplexed over that single connection. QUIC also enables network path migration without losing the connection, which is a handy feature to reduce streaming interruptions when moving between networks. Although H3 inherits most of the features of H2, it provides additional features, too, due to the use of QUIC as opposed to TCP. One such feature is the zero round-trip time (0-RTT) connection setup. While H2 wastes some RTTs for the three-way handshake due to TCP connections, H3 takes advantage of Transport Layer Security (TLS) 1.3 (RFC 8446) directly integrated into QUIC to save at least one RTT for connection setup. H3 is also free from the HoL blocking issues inherent to TCP. Different requests can be processed simultaneously with stream multiplexing by putting each pair of HTTP requests and responses in an individual stream, where a stream is an independent sequence of data delivered between the server and client.

The server push feature in H2 and H3 enables the client to receive multiple responses from the server by sending a single HTTP GET request. This may be helpful in instances where the server knows which responses are necessary for the client to process the corresponding request completely. This feature helps the client save multiple RTTs compared to non-pipelined HTTP/1.1, where a new request is sent only if the response of the previous request has been fully received. HTTP/1.1 offers the pipelining feature (a feature that many browsers have disabled by default due to the buggy proxy implementations and other concerns) that allows many requests to be delivered over a single TCP connection without waiting for each response. However, it needs more requests than H2 and H3 to obtain the same responses. Fig. 4 shows a comparison of the server push feature in H2 and H3 to HTTP/1.1 (non-)pipelining.

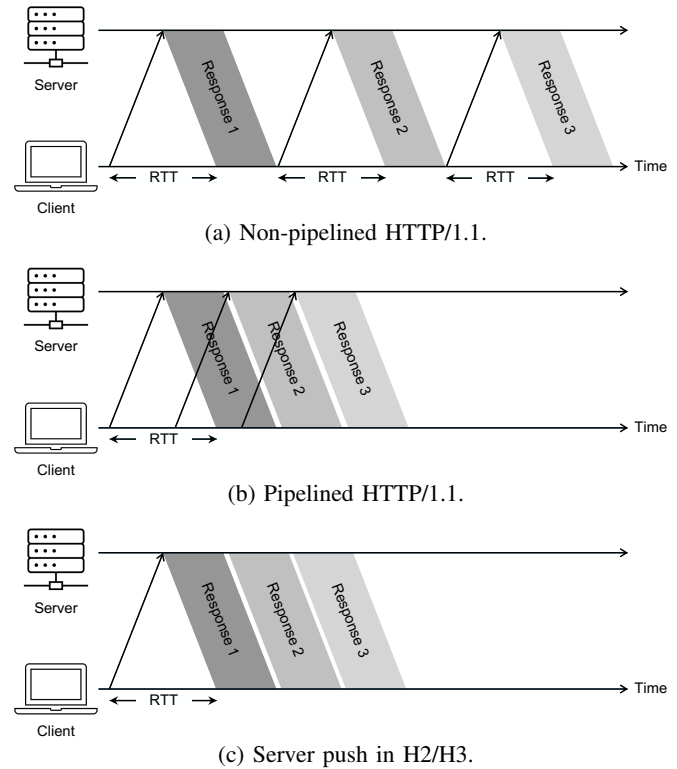

Fig. 4: The server push feature in H2 and H3 compared to HTTP/1.1. (a) In non-pipelined HTTP/1.1, the pairs of requests and responses are sent and received one after another, which uses one RTT for each pair. (b) When pipelining is enabled for HTTP/1.1, some RTTs can be saved, but the number of requests remains the same. (c) H2 and H3's server push feature eliminates the need for multiple RTTs with only one request.

The server push feature can be used to reduce the latency in live media streaming in addition to reducing the number of requests, especially when using short-duration segments. For example, Wei *et al.* [221] developed an H2 push-based architecture for LLL streaming for DASH. The architecture considered three push strategies: no-push, all-push and $k$-push. In no-push, the client sent a request for each segment, and the server responded to each request with the corresponding segment. In all-push, the client only sent one request for the entire live session. Upon receiving the request, the server pushed all the segments sequentially as soon as they became available. In $k$-push, the client sent one request for $k$ segments, and the server responded by pushing $k$ segments consecutively. Similarly, Xu *et al.* [225] designed a QoE-driven H2 $k$-push solution for LLL streaming. The DASH player was responsible for setting the parameter $k$, selecting the bitrate level and issuing an HTTP request, while the server would push $k$ corresponding segments at the selected bitrate in a batch. The $k$ parameter was determined based on a probabilistic buffer model considering buffer underflow or overflow cases. At each push cycle, the problem of finding a suitable *k* parameter and ABR decision were formulated as a multi-objective optimization problem (Pareto optimal), which was solved via the Nash bargaining solution.

In yet another study, Le *et al.* [127] presented an ABR scheme for LLL streaming over a mobile network using H2's server push and stream termination features. The server kept pushing the two-second video segments without requiring a request, and the client could perform ABR decisions even when a segment was being downloaded. If a segment could not be supported by the network, the client used stream termination to cancel this segment and request its lower-bitrate version. Similarly, van der Hooft *et al.* [208] described an H2 push-based solution for LLL streaming with super-short segments (less than one second). The authors analyzed the encoding overhead for super-short segments and discussed the best trade-off between responsiveness, overhead and segment duration. Benyahia *et al.* [226] developed a frame-discarding algorithm to meet the one-second latency constraint and alleviate rebuffering events in LLL streaming under fluctuating network conditions. The proposed solution requested frames of a segment individually through H2's multiple streams feature and discarded the least meaningful video frames using H2's stream resetting feature. The use of the server push feature for LLL streaming is specified with examples in one of the parts of the DASH standard family [106]. However, the market uptake so far has been rather low compared to other approaches.

Among the multiple streams over a single connection, the client can indicate its priority for a stream over the others by the stream priority feature. If the stream priority is not indicated, these streams share the connection's throughput equally. This feature exists in both H2 and H3. The IETF recently published an extensible prioritization scheme in RFC 9218. New protocols can intelligently use this prioritization scheme to achieve timely delivery [71].

HAS has become the de facto technology to scale media delivery and accounts for the majority of Internet media traffic today. This is primarily because HTTP scales well due to its stateless nature, features well-understood naming/addressing and authentication/authorization schemes, provides easy traversal through all kinds of middleboxes and leverages the existing, cheap HTTP caching infrastructure. In HAS, clients imitate streaming via short downloads as they *pull* the media (in segments) from a standard HTTP server that hosts the media content. The media is offered in multiple representations to provide clients with adaptation capability based on network conditions and other factors. This simple idea scales exceptionally well and effectively improves user experience compared to server-centric, stateful, heavyweight protocols, such as Real-time Streaming Protocol (RTSP)[1]. While there have been several different HAS methods developed and deployed over the past two decades, only two remain: DASH and HLS. For the obsoleted methods and other historical data, refer to [40].

In DASH and HLS, the server is typically a simple HTTP server that serves mixed live and on-demand content, each stored in fixed (or occasionally variable)-duration segments. Segments are commonly two to ten seconds long and generated from different DASH representations (or HLS variant streams) encoded using a bitrate ladder. The addressing information for the segments is provided in a manifest file, also called a media presentation description (MPD) for DASH or master playlist (M3U8) for HLS. The manifest contains other details such as the codecs used for each representation, metadata, information about the encryption keys and licensing server(s), and the relationships among various tracks of the same content (*e.g.*, video, audio and subtitles). Once an HTTP connection is established, the DASH/HLS client first fetches the manifest file and then starts requesting media segments sequentially from the server. During the session, the client runs a rate-adaptation algorithm (also called an ABR scheme or logic) that monitors the throughput for every segment download and the (playback) buffer occupancy, among other things, to dynamically select the segment to be downloaded next. While most rate-adaptation algorithms try to select the segments encoded at the highest bitrate they can afford to download without risking a playback rebuffer, developing more sophisticated algorithms has been an important research topic in the past decade. The work in [40] provides a survey of ABR schemes, including various client-based, server-based, network-assisted and hybrid rate-adaptation solutions that use buffer size, available bandwidth and other input metrics in their logic. In Section VII-B, we also discuss how ABR schemes can be better designed for LLL streaming. Further details of DASH and HLS are discussed below.

- **Dynamic Adaptive Streaming over HTTP (DASH)** [109] is an international standard developed by many experts from the industry and academia working together under the umbrella of MPEG (under ISO/IEC JTC1 SC29). It is an open specification defining media segment and manifest formats for adaptive streaming over HTTP. In addition to these formats, it shows how DASH can support streaming monetization through content protection and ad insertion. Fig. 5 shows the essential components of DASH where boxes in blue are specified in the DASH standard, and red ones are either specified by other standards or left unspecified to spur innovation. The official work in MPEG started in 2010, and the first edition of the core part of the standard (ISO/IEC 23009-1) was published in 2012. Later, this part's second, third and fourth editions were published in 2014, 2019 and 2019, respectively, with several amendments and corrigenda published in between. The latest (fifth) edition was published in the second half of 2022 [109]. At the writing of this survey, the DASH standard family has eight other parts (nine parts in total) that complement the core specification in several ways to address different market needs.
- **HTTP Live Streaming (HLS)** [16] is a specification

---

[1] Initially specified in 1998 in RFC 2326, RTSP has been a popular streaming protocol for about a decade. However, its popularity plummeted since then. Later, RTSP 2.0 (RFC 7826) was published, but there has not been a notable implementation or deployment.

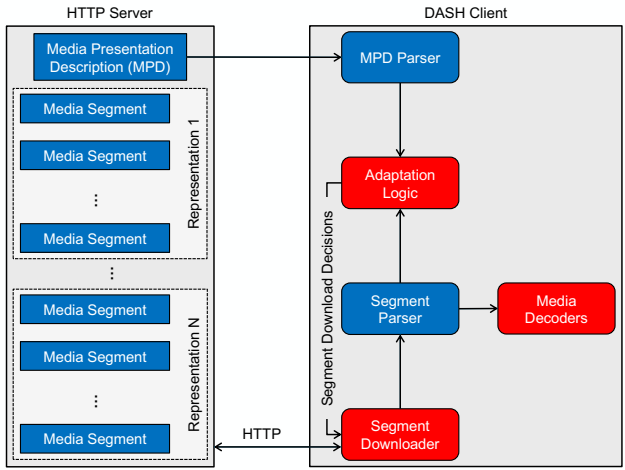

Fig. 5: Essential components of DASH. Boxes in blue are specified in the DASH standard and boxes in red are either specified by other standards or left unspecified.

developed and controlled by Apple to define its own media segment and manifest formats. HLS supports streaming both live and on-demand content. At its inception, HLS targeted mobile devices and was exclusive to iOS running on iPhones, and gradually made its way to all Apple devices. Today, beyond the Apple ecosystem, Android devices and most smart TV platforms also support HLS. The primary HLS specification has been published as RFC 8216. As Apple adds new features, they are documented in a follow-up Internet-Draft [174].

*2) Peer-to-Peer (P2P) Delivery:* While P2P systems have a long history in many application areas, the architecture became popular with the file-sharing system Napster, originally released in 1999. P2P systems collectively work on a joint task in a distributed manner. One of the key characteristics of P2P systems is that each peer (node) functions as a server to a subset of its connected peers while acting as a client to another subset of its connected peers. In P2P live streaming systems, the set of peers (independent streaming agents) create an overlay network and organize themselves to help distribute live content to all peers in a timely manner. The performance of P2P live streaming systems is usually measured by the delay-loss curve, *i.e.*, what is the required playback delay to achieve an acceptable media quality, expressed as the segment miss-ratio at the peers (where segments are the blocks of media exchanged between the peers). The media quality is determined by many factors, such as the overlay construction algorithm, the forwarding algorithm, the loss-recovery techniques in the underlying network, the number of peers in the overlay network and their bandwidth distribution, the contribution of the peers with their resources and the viewing behavior of the peers (*i.e.*, viewer arrivals and departures, or viewer churn).

P2P media delivery systems can be decomposed into two aspects that work in parallel: (*i*) overlay construction and maintenance, and (*ii*) segment forwarding. Depending on the design choices for both parts, a P2P architecture can be classified into being *tree-based* (single and multiple), *mesh-based* or *hybrid* (tree and mesh).

In a tree-based system, *e.g.*, end-system multicast (ESM) [56] and SplitStream [51], the peers are organized in layers that consist of rooted distribution trees, where each layer has a *seeder* (usually representing the root or parent of the (sub-)tree) and a set of *leechers*. The seeders deliver the segments to the leechers while organizing the join/leave processes of their leechers. In general, each seeder in the tree can have as many leechers as its capacity relative to the streaming rate allows. The tree construction and seeders-leechers relationship can be determined by factors like the latency between peers, available network capacity or underlying physical topology. P2P tree-based systems generally utilize the available network capacity well. However, they are vulnerable to sudden peer departures (churn) and tree maintenance overhead problems [135].

To address these problems, mesh-based systems, *e.g.*, CoolStreaming [235], were introduced. They focus on creating an overlay where every node is connected to *multiple* peers that act as seeders (and leechers) concurrently. This changes the topology from a loop-free tree to a mesh and makes it resilient by design, *i.e.*, a node departure or failure no longer results in a disconnected sub-tree. A peer receives data from multiple neighbors, making the overlay construction simple and fast. However, sophisticated forwarding algorithms are needed at the peers to address looping and avoid data duplication when forwarding. One approach to make data distribution robust is to employ *network coding* [58], [82], [142], [164]. In network coding, data is divided and mathematically encoded into multiple fragments such that the original data can be reconstructed when receiving a subset of the fragments. Mathematical encodings that have been used for peer-to-peer distribution are, for example, *fountain* and *raptor codes*, which fall into the class of *rateless codes* [47], [138], [182]. In mesh-based systems, the end-to-end delay may be more difficult to be bounded because different data blocks can travel over different paths. Furthermore, additional overhead is generated due to the exchange of control messages (*e.g.*, buffer maps) between peers.

Another example of such mesh-based P2P systems is BitTorrent [46], which was developed more than two decades ago for efficient and scalable replication and distribution of large amounts of static data. Due to its features and popularity, BitTorrent has been used for legal distribution and illegal sharing of movies, TV shows and songs. It was first extended to support the streaming of pre-recorded content [55], [64], [213], [220] and then live content [152], [178], [198].

To leverage the advantages of tree and mesh-based systems, hybrid systems, *e.g.*, MultiPeerCast [137] and Layeredcast [153], were introduced. They combine the robustness and simplicity of a mesh overlay construction with the efficient forwarding over a fixed structure that tree-based systems provide. Nonetheless, hybrid systems struggle with the complexity of the trade-off between stability and scalability [100].

WebRTC [215] is an API defined by the World Wide Web Consortium (W3C). In addition to direct communication, WebRTC data channels (RTCDataChannel API) also support P2P transport as a protocol without needing a browser plugin, extension or other installation. Viewers connected to a Website can form a distributed, decentralized browser-to-browser network for low-latency and encrypted data transfer. This feature enables the use of WebRTC for streaming live media [24], [149], [170], [209]. This topic is further explored in Section VI-A.

### E. Challenges in Protocols for Live Streaming

The different protocols surveyed in this section show the current landscape of live media streaming. Table IV highlights a high-level comparison. The primary challenges of these protocols can be categorized as follows:

- *Latency*: Ensuring minimal delay from capture to playback.
- *Scalability*: Supporting a growing number of concurrent viewers without compromising QoE.
- *Bandwidth Management and Reliability:* Being capable of utilizing varying available network resources efficiently to provide a good and consistent QoE for the viewers.
- *Security:* Possessing end-to-end security mechanisms to protect live media assets against various types of attacks.
- *Interoperability and Ease of Deployment:* Featuring interoperability with other protocols used in the contribution, ingest and distribution stages and easier integration with existing network infrastructure, requiring minimal engineering adjustments.

In response to these challenges, new protocols, such as SRT, RIST and WHIP, are being developed. Further investigations into these and other technologies are needed to achieve performant and reliable LLL streaming systems. Also, a deep understanding of the performance, limitations and design implications of different protocols is critical to make good decisions regarding the deployment of LLL systems.

## IV. Latency Sources

In this section, we dissect the E2E latency into three serial parts for a more organized examination: media content preparation, delivery and consumption latency. We provide an in-depth analysis of the components contributing to the latency for each part. Then, we highlight improvements for each component to meet the LLL streaming requirements.

The Moon landing in 1969 was one of the most significant events broadcast on TV in history, reaching an estimated 650 million viewers. The image and sound signals were transmitted via a lightweight antenna on the top of the lander over 250 thousand miles back to Earth. The broadcast latency from the face of the Moon to Earth was approximately three seconds. Half a century later, in 2019, Super Bowl LIII was broadcast using 115 cameras (including 8K ones), AR-powered graphics and many other innovations. Despite all the technological advancements, the latency was higher than 45 seconds from the stadium to a streaming device located only five miles away.

Today, audiences are picky and demand high-quality content and smooth delivery no matter what devices they use or where they are. They also prefer watching certain content with low latency to keep the excitement and engagement high. While scaling live streaming services is already operationally challenging, offering a low-latency live streaming service is even more so and puts significant stress on the content providers since quality-related problems may quickly become more severe due to the tighter constraints in the system.

The E2E media delivery workflow is complex, comprising a range of components from a live source to media players, each contributing to the E2E latency at varying degrees. In order to effectively reduce latency, the entire media delivery pipeline must be streamlined. One poorly configured or malfunctioning component can harm the whole effort [79]. In this section, we discuss in detail the three main parts of E2E latency and the different components that contribute to each part. The main parts are identified as media content preparation latency (Section IV-A), delivery latency (Section IV-B) and consumption latency (Section IV-C). An overview of the latency parts and their contributing components is also shown in Fig. 6.

Latency measurements can be achieved in many different ways [195], including (*i*) capturing a live clock and comparing it to the time of the playback, (*ii*) embedding a timestamp into the video frames, (*iii*) frame tagging/fingerprinting (*e.g.*, using SMPTE ST 2064 [190]), and (*iv*) inserting specific metadata into the workflow (*e.g.*, using picture timing supplemental enhancement information (SEI) messages [107], EXT-X-PROGRAM-DATE-TIME tags, `ProducerReferenceTime` element and `prft` boxes [108]).

### A. Media Content Preparation Latency (MCPL)

The live content (or content from a linear source) must first be acquired and then prepared for OTT distribution. Beyond the acquisition latency, which is usually negligible, the MCPL comprises latency introduced in encoding, packaging and ingest, which are described below.

1) The (contribution and distribution) encoding latency is often the largest contributor to the MCPL[2]. It is impacted by many factors, including the choice of codec, content complexity, target quality and encoding parameters (*e.g.*, bitrate ladder, and group-of-pictures (GoP) and segment duration).
2) The packaging latency is introduced because of the encapsulation of the encoded stream into one or more delivery formats. An optional encryption step may further increase this latency.
3) The ingest latency is introduced when uploading the packaged media segments and the associated manifests to an origin server.

Next, we discuss the details of each latency component.

---

[2]Since video encoding is significantly more complex and takes a much longer time than audio encoding, we mainly focus on the latency due to video encoding in this section.

TABLE IV: Pros and cons for some of the protocols surveyed in this section.

| Protocol | Pros | Cons |
|---|---|---|
| RTMP | - Low latency.<br>- Flexibility in streaming. | - Not supported by HTML5.<br>- Vulnerable to bandwidth issues. |
| RTSP | - Suitable for segmented streaming.<br>- Customizable for various applications. | - Incompatibility with HTTP.<br>- Less popular in modern streaming architectures. |
| DASH/HLS | - Wide compatibility across devices.<br>- Scalability for large audiences. | - Higher latency compared to other protocols. |
| WebRTC | - Real-time communication with minimal latency.<br>- Native support by the browsers. | - Limited scalability for large broadcasts.<br>- Complex implementation. |
| SRT | - Secure and reliable transport.<br>- Low-latency performance. | - Limited support and adoption in the industry. |
| MOQ | - Promising low-latency streaming over QUIC.<br>- Improved security. | - Emerging technology with limited real-world deployment data. |

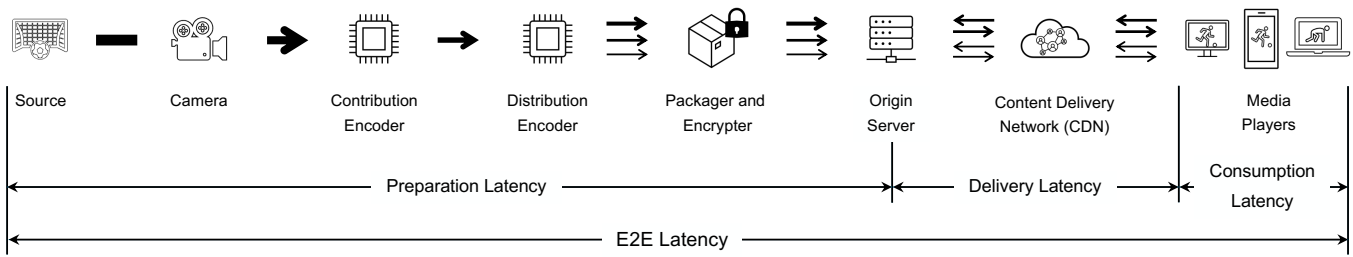

Fig. 6: E2E live streaming architecture (adapted from Fig. 3) showing media content preparation, delivery and consumption latency and the components contributing to each latency.

*1) Encoding Latency:* According to [193], there are three main objectives that every video codec should meet for better streaming performance: (*i*) high compression ratio (to reduce bandwidth and storage costs) while maintaining (*ii*) high visual quality, and (*iii*) low complexity and fast processing (which directly impacts latency). Selecting the right codec is not sufficient to reduce latency, as we also need to pick a set of suitable parameters for the encoding. Such parameters include the number of encoding bitrate levels (also called the bitrate ladder or encoding profiles), and GoP and segment duration. There are often trade-offs in the codec and encoding parameters that impact latency, and we discuss the notable contributing factors below.

**Impact of Bitrate Ladder Selection.** The bitrate ladder specifies the bitrate-resolution pairs at which the video is encoded and can be streamed. A bitrate ladder with more granular or customized steps may increase the encoding complexity, and thus, the latency and cost [102], [118]. However, it may also give streaming clients more options in rate adaptation based on factors such as device resolution and varying network conditions. Evidently, selecting an appropriate bitrate ladder for different genres and different videos is an important yet challenging task.

One of the earliest developments in this area is per-title encoding (PTE) [3]. PTE customizes the bitrate ladder for each video asset based on content awareness (*e.g.*, content type and complexity) and context awareness (*e.g.*, network conditions, device capabilities and streaming service requirements). Netflix showed that PTE could provide up to 20% reduction in bitrate at the same visual quality [3], although its complexity, and hence, its impact on latency, could be further improved. In this regard, machine learning (ML) approaches [86], [101], [184] were developed to automate the PTE process more efficiently, thereby reducing its processing time and cost.

**Impact of GoP and Segment Duration.** GoP duration determines the interval between key frames in an encoded video. A key frame is where the decoder can start decoding from. In order to support seamless switching between different encodings for rate adaptation, media segments always start with a key frame. In other words, each media segment must have at least one GoP. This is often the case for media segments that are two seconds long or shorter, while longer media segments tend to have two or more GoPs, and hence, more key frames. We note that key frames are substantially larger in size than other predictively-encoded frames. With longer media segments, rate-adaptation granularity worsens, and sudden bandwidth drops may lead to more frequent rebufferings.

If the low-latency extensions are not used, the E2E latency becomes directly proportional to the segment duration, as exemplified in Section IV-C. Using shorter segments (thus, shorter GoPs) may help reduce the E2E latency, however, using more frequent key frames reduces video encoding efficiency and is generally not advisable. Also, if the download speed of the streaming client is not significantly larger than the encoding bitrate or the network delay between the client and server is not negligible, the startup delay also increases with longer segments due to their size increase. If the segment duration were kept the same, but the GoPs were shortened, each segment would now have more than one key frame, which increases the segment size, download time, and possibly the E2E latency. This effect becomes more significant with longer segments (*e.g.*, > six seconds) or when the GoP duration falls below two seconds, as shown in [42], [199].

*2) Packaging Latency:* After the source media is encoded into the different representations, the packaging process begins to package the content into various delivery formats, such as DASH and HLS. One important task here, for example, is to create the manifests and media segments that conform to the respective formats. While packaging is generally less computationally intensive than encoding, the need to cater to and produce multiple delivery formats may add considerable latency to the workflow. To mitigate this, [8] proposed a solution that first packages the content in a single format and, when necessary (*e.g.*, when a player does not support the currently available format), repackages the segments dynamically at the edge. Such solutions may be effective if there is a format far more popular than the others, as this increases the utility of the first packaging. Another advantage of such solutions is that, as CDNs will only need to deliver a single format, there will be better cache and bandwidth utilization, which may further reduce latency. We note that many other implementation-specific factors may also contribute to the packaging latency, such as the various packaging parameters and available computing resources.

In recent years, several notable advancements in streaming standards and protocols have also sought to reduce packaging latency. For instance, the CMAF standard was developed in an effort to converge and simplify the various delivery formats, which is discussed in greater detail in Section V-A. Also, DASH and HLS have introduced their low-latency protocol variants (namely, LL-DASH and LL-HLS) that provide new features to enable better performance in LLL streaming scenarios, and we discuss these protocols in greater detail in Sections V-B1 and V-B2.

The packaging process also typically includes the encryption of streams into various DRM-protected formats such as Widevine [85], PlayReady [150] and FairPlay [20]. Similarly, it is important to design systems that can run such operations with minimal latency overhead. Due to the rise of new security threats, additional encryption may also be needed, *e.g.*, at the transport layer, to protect the media and manifest data against fingerprinting or tampering attacks [88], [111]. For instance, [176] showed how transport layer data, *i.e.*, TCP headers, could be used to gain information about content streamed over Netflix even though the transmission was HTTPS-encrypted. Streaming service providers need to constantly seek a balance between such privacy and security concerns, and the additional latency introduced by these operations.

*3) Ingest Latency:* Ingest is the process of transmitting packaged content to the origin server, which stores and readies them for retrieval (usually by CDN servers or media players themselves). Section III-C discussed several media ingest protocols, which are designed with different considerations, and may hence, contribute differently to the E2E latency. For instance, the likes of WebRTC and RUSH are designed for live media ingest and tend to prioritize low latency over other factors, such as reliability. Studies have also been done to analyze such performance differences empirically. The work in [123] compared the trade-offs between the WebRTC and SRT protocols and found that while SRT could provide better quality, the latency experienced in its experiments was up to four seconds (as compared to the sub-second latency achieved with WebRTC).

Other tasks may also be introduced at the ingest server, and their contributions to latency should also be carefully inspected and minimized. One example is LiveNAS [121], which is designed as a live streaming ingest framework to improve video quality through super-resolution and online learning. To minimize the latency introduced by these tasks, LiveNAS uses multiple GPUs to produce inference results in the shortest time possible. We also note that the diverse (and ever-changing) options in ingest protocols and tools have made designing, implementing and updating ingest workflows a complicated task. To simplify this, [156] designed Nagare Media Ingest, which is a flexible multi-component ingest server that separates the ingest tasks into changeable components. Such solutions are important in LLL streaming as the space is actively evolving and new technologies may be developed over time. Hence, systems should seek to be adaptable and constantly update to better alternatives that could further reduce latency.

### B. Media Content Delivery Latency (MCDL)

The MCDL is primarily due to two main tasks: CDN distribution and last-mile delivery.

1) The CDN distribution latency is introduced when propagating the media segments through different caches of the CDN cloud.
2) The last-mile delivery latency differs depending on which access network the viewer is connected to during streaming. Also, additional latency could be introduced depending on the geographical location and position of the closest CDN end-point.

Below, we discuss the details of each latency component.

*1) CDN Distribution Latency:* In order to handle the large count of requests coming from media players, having CDNs in the delivery workflow is an absolute necessity. CDN clouds are responsible for distributing all kinds of objects (not only media segments and manifests but also ad assets and player codes/scripts) through caches, bringing these objects closer to the edge and serving millions of media players around the globe. CDNs reduce the time it takes to download these objects.

An OTT service provider may offer its content through more than one CDN provider to improve the reliability and reach of its service. This is referred to as multi-CDN distribution [19], [76], [185], [186]. The multi-CDN service is a way to combine the strengths of different CDN providers while minimizing the weaknesses of each (the risk of degraded performance and outages). A single CDN provider, for example, might have great coverage in Asia, while another CDN provider might have great coverage in North America and Europe. When one uses a multi-CDN service [204], the traffic can be routed to different CDNs in a way that makes the most sense in terms of real-time performance and availability. Having alternatives allows service providers to protect their viewers against negative user experience impact caused by a single CDN [43], [63]. Although CDNs bring many benefits, they could also become a bottleneck and substantially increase the E2E latency. Therefore, selecting a CDN provider that is good at LLL streaming is an important consideration.

**Impact of CDN Provider Selection.** To reduce CDN distribution latency, one key objective is to select CDN provider(s) that can (*i*) provide suitable CDN server locations with minimized distance to viewers and support the low-latency key enablers such as chunked encoding/packaging and delivery (further elaborated on in Section V), (*ii*) serve live content off the memory rather than disk, and (*iii*) minimize cache misses. Each CDN provider has a set of points of presence (PoPs) strategically located and responsible for serving viewers in their geographic vicinity. Their primary function is to reduce the RTT by bringing the content closer to viewers. Each CDN PoP typically contains numerous caching servers responsible for storing and delivering media assets. Each server holds multiple storage drives and high amounts of memory. Inside these servers, objects are stored on solid-state drives (SSDs), spinning disks and in the memory, with the more commonly used files hosted on the more speedy mediums. Being the fastest of the three, memory is typically used to store the most frequently accessed items and for LLL services.

Moving on to the objective of minimizing cache misses, a cache miss happens when the requested object is not available at the cache and the request is forwarded upstream to another cache (and eventually to the origin server). To meet this objective, designing effective caching policies and prefetching decisions is essential. Caching policies are the set of rules that CDNs apply to store the objects, and prefetching decisions are the set of techniques that CDNs use to prefetch objects from the origin server (before there is a request for them) to boost cache hits.

The caching policies and prefetching decisions become important once the live streams are made available in the CDN cloud, as the media players would be able to access CDN caches close to their geographical location instead of having to reach the origin server for faster delivery. However, in LLL streaming, the E2E latency is kept low, which makes caching policies and prefetching decisions a complicated task. Therefore, it is not uncommon that viewers joining a live stream experience occasional cache misses. For example, Meta [146] reported that while broadcasting a live event to millions of viewers in 2015, 1.8% of the requests were routed to the origin server due to cache misses, resulting in many media players suffering from poor performance.

In LLL streaming, media players are generally aggressive in their requests and may ask for a media segment that has not been generated yet, resulting in an HTTP 404 error. Each CDN has a unique default time-to-live (TTL) value for caching these 404 errors, and typically, these values do not perform well for LLL streaming. There have been several proposed solutions to tackle these issues [14], such as adjusting these parameters dynamically, disabling the artificial buffers in CDNs and ensuring the TTL of the DASH manifest or HLS playlist is shorter than or equal to the DASH manifest update interval or HLS segmentation interval. Other strategies have also been designed to mitigate the impact of cache misses.

When a cache miss occurs, the CDN server sends a cache fill request upstream. In the case of multiple cache misses on the same object by multiple media players (*e.g.*, a common scenario during the start of a live stream), the CDN servers should send only one cache fill request to minimize the load on the origin server, a feature known as request coalescing. This can be done using a cache-fill-wait-time approach that pools the set of requests for the same object and sends only one request upstream [74]. Once the CDN cache starts loading, it forwards the partially filled cache to the various players via partial cache sharing. Another similar approach that employs the idea of request pooling in the CDN is the Pending Interest Table (PIT) [54]. To further improve the cache fill process, CDN servers may also use sub-request spawning to decouple the process from the requesting media player so that it is not limited to the media player's capabilities (*e.g.*, connection speed). Also, when a popular stream is detected within a PoP region, CDNs can quickly replicate the content across the nearby caches to handle the anticipated requests from other media players. In such cases, the replication delay should be as small as possible. For example, [74] designed a solution that could reduce replication speed from five seconds to approximately one to two seconds.

**Impact of Flash Crowds.** Another prominent challenge that CDNs need to consider and handle well is when there is a sudden and significant increase in the number of media players. This phenomenon is known as *flash crowd* and frequently occurs when many viewers join a live stream event at once – such as when viewers are hungry to catch the kickoff or overtime action in a football game [222]. For instance, Verizon [74] investigated the rapid jumps in the number of viewers by observing more than 100,000 live sports events, of which, during the NBA finals, the viewer population grew rapidly from almost a few hundred at tipoff to a peak of 2.04 million viewers in the third quarter. In the beginning, the number of viewers was less than 10,000, but it jumped to over one million in the first hour, and another one-and-a-half million viewers joined after half-time. Such rapid growth puts considerable pressure on CDNs, which may result in poor user experience if the CDNs are not designed to deal with the rapid scaling of viewers. In other words, CDNs need to carefully design their interactions with both the origin server (prefetching decisions) and the media players (caching policies) while scaling for large pools of viewing audiences that may come all at once. CDNs should aim to minimize the load on the origin server and keep latency to the lowest possible. One potential solution [74], [81] is to use an origin shield between the origin server and CDNs. The shield acts as an intermediate caching proxy that helps alleviate flash crowd issues, as depicted in Fig. 7. The origin shield can be deployed

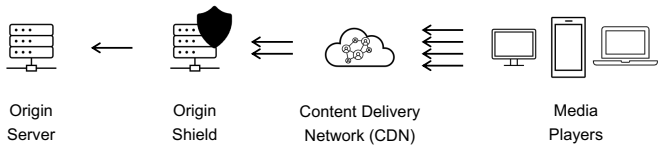

Fig. 7: Deployment of the origin shield.

in one of the existing PoPs that manages all the requests from various CDNs in the same region.

*2) Last-Mile Delivery Latency:* Last-mile delivery refers to propagating media content from CDNs to viewers' devices via an access network – usually an institution, home or cellular network provided by an Internet service provider (ISP). In recent years, there has been a significant increase in mobile data usage in cellular access networks due to the large consumption of entertainment content [111]. Such high-bandwidth video streams can strain the network and cause congestion, especially during peak hours. Fortunately, advancements in 5G networks have made LLL streaming over mobile more plausible than before, as real-world download speeds have increased from approximately 24 Mbps (4G) to 165 Mbps (5G) [136].

Increased speeds, for instance, allow the startup delay of a typical 4K video to be reduced from a few seconds to a second or less. Similarly, they also help reduce the last-mile latency. However, the move from wired to wireless connections in access networks (*e.g.*, WiFi, 5G, LTE) has also posed other challenges to ISPs and streaming service providers, such as having wider and more rapid variations in link capacity. In real-world environments, wireless links often face capacity fluctuations, especially when the viewer's device changes location or there is interference in the transmission (*e.g.*, interference from home appliances like microwave ovens). Such a behavior can lead to rebuffering events, and hence, increased latency if not properly taken care of by the media player.

Network providers can also look into network-specific enhancements to further reduce latency. In the case of 5G non-public networks (NPN), for instance, studies have been conducted to investigate latency-contributing factors unique to the technology and recommend targeted interventions within these private networks. One such study [216] was conducted for a 5G NPN used in low-latency broadcast news production. The study found several causes of latency related to 5G technology, such as factors relating to its capacity scheduling and retransmission mechanisms. To this end, the study proposes several network configuration patterns to cater to different latency targets (usually with a trade-off in capacity) and use cases. Another well-studied category of enablers for 5G NPN is the use of network slicing and other software-defined networking (SDN) techniques [171], [231]. Among other things, SDN allows network connections to be prioritized and allocated resources dynamically to meet intricate low latency targets. As this survey focuses on technologies in the application layer, we refer interested readers to the included references for further reading.

Technologies within the media delivery space evolve continuously and rapidly, and other developments may also contribute to the solutions in this space. Some examples include advancements in mobile communication networks (5G mmWAVE, 6G and beyond) [2], [113], [120], [179] and multi-access mobile edge computing [80], [116], [167], [197].

*C. Media Content Consumption Latency (MCCL)*

MCCL is mainly introduced because of the many media player tasks such as rate adaptation, buffering, playhead positioning, license/key fetching, decryption, decoder priming, decoding and rendering. Media players are typically designed to ensure smooth playback, which is achieved by buffering several seconds of media. In LLL streaming, buffering is severely limited due to low-latency constraints. The media player is one of the more influential components contributing to latency, which may add up to 50% of the E2E latency if it is not tailored for LLL streaming. Hence, adapting the player for LLL streaming can go a long way [125].

Consider Fig. 8, which demonstrates a live streaming scenario where the content is packaged into four-second segments, and the numbers inside the boxes indicate the segment numbers. After downloading the manifest, a media player joins the live session 18 seconds into the content. There are three common strategies the media player can adopt (each of which contributes differently to the latency):

- The media player requires at least $m$ segments to be downloaded fully before the playback can start. For the native (HLS) media player on iOS/iPadOS/tvOS platforms, $m$ equals three. The media player checks the playlist and determines that segments #1 through #4 are available so far. Consequently, it requests segments #2, #3 and #4. Assuming the segment fetching time is negligible (meaning that the RTT between the media player and the server is small and the available bandwidth is plenty), the media player can start the playback quickly but with a latency of 14 seconds.
- The media player fetches the last fully available segment. In this case, the latency drops to six seconds.
- The media player prudently defers sending the request for segment #4 at the time of joining, waits for two seconds and then sends the request for segment #5. This way, the media player achieves the lowest latency possible, at four seconds, although the startup delay increases by two seconds.

Downloading the media segments is not sufficient to start the decoding and rendering. The media player must also communicate with the licensing server, fetch the keys to perform decryption, and then prime the video and audio decoders. Any extra delay in this process may negatively affect the latency as well as the startup delay. Beyond this, there are other new challenges (in addition to those discussed above) for the media player where modules like bandwidth measurement, playback speed control, (playback) buffer management and bitrate selection scheme become non-trivial. These modules should be re-visited and re-designed to support streaming at low latency. We continue this discussion further in Section VII.

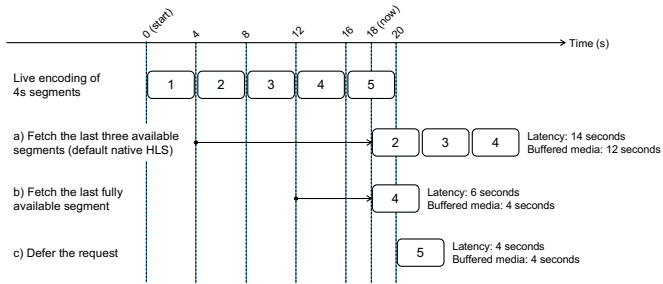

Fig. 8: Different playback strategies for segmented content, where each segment is four seconds long. Segment fetching time is assumed to be negligible.

## V. HTTP-Based Low-Latency Live Streaming

This section discusses the key technologies driving HTTP-based LLL streaming: CMAF, chunked encoding/packaging and chunked delivery. After the preliminaries, we dive deeper into the low-latency extensions for DASH and HLS, and discuss implementation-specific details. We then show how a common chunked-packaged media can be efficiently served using the low-latency extensions for both DASH and HLS. We wrap up the section by discussing relatively new HTTP-based protocols for LLL streaming, called High Efficiency Streaming Protocol (HESP) and Low-Latency Low-Delay Extensions for DASH (L3D-DASH).

### A. Key Enablers

*1) Common Media Application Format (CMAF):* In 2016, several technology providers and vendors joined forces to develop an extensible encoding standard to unify the streaming media format and eliminate the proliferation of alternatively packaged content. This standardization effort resulted in the Common Media Application Format (CMAF) [110] specification [28]. CMAF uses the ISO Base Media File Format (ISO BMFF, ISO/IEC 14496-12) for the common core media segments and allows the use of any format for the manifest during delivery.

Fig. 9 depicts the data structure for a CMAF fragment, which is the media object encoded and decoded independently in the CMAF hypothetical application model. The smallest CMAF media object is a media sample, which is media data associated with a single decode start time and duration, such as a video or audio frame. A CMAF fragment typically consists of one MovieFragmentBox (`moof`) and MediaDataBox (`mdat`) pair but can also contain more than one of these pairs. If the CMAF fragment contains multiple such pairs, each pair is called a CMAF chunk and contains a consecutive subset of the CMAF fragment's media samples. A CMAF fragment is independently decodable and randomly accessible, and seamless switching across different tracks can only happen at fragment boundaries. In combination with the associated CMAF header, a CMAF fragment must contain the metadata necessary to decrypt, decode and display it. On the other hand, CMAF chunks are not necessarily self-decodable, *i.e.*, not every chunk can be used as an access point.

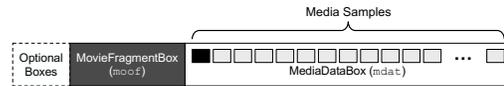

Fig. 9: Data structure of a CMAF fragment containing a single CMAF chunk.

CMAF chunks are particularly critical for low latency, as described below, since a CMAF chunk can be delivered as soon as it is encoded/packaged without waiting for the encoding/packaging of the subsequent chunks in the same CMAF fragment. CMAF chunks cannot reduce the latency themselves. To achieve a low E2E latency, they need to be delivered with an appropriate method, and this is where chunked delivery comes into play.

The first chunk in a CMAF fragment is signaled by a SegmentTypeBox (`styp`) that contains both the `cmff` (indicating the start of a CMAF fragment) and `cmfl` (indicating the start of a CMAF chunk) brands. However, the other chunks in the same CMAF fragment are signaled by an `styp` that contains only the `cmfl` brand.

One or more CMAF fragments can be packaged into a larger and addressable media object called a CMAF segment, depicted in Fig. 10. CMAF segments are typically several seconds in duration. Unless CMAF chunks were supported and used, one would not prefer long segments as they would increase the E2E latency and require the streaming client to switch tracks at longer intervals (*i.e.*, at a reduced rate-adaptation granularity).

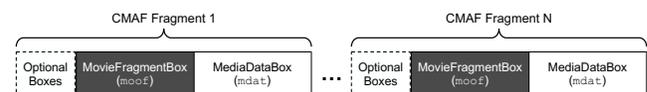

Fig. 10: Data structure of a CMAF segment.

A CMAF track is a logical, continuous sequence of one or more CMAF fragments in presentation order, conforming to a CMAF media profile and the associated CMAF header. The CMAF header contains a FileTypeBox (`ftyp`) and a MovieBox (`moov`), sufficient to process and present all the CMAF fragments in the CMAF track. CMAF tracks may exist temporarily while streaming or be permanently stored as CMAF track files. The data structure of a CMAF track is shown in Fig. 11.

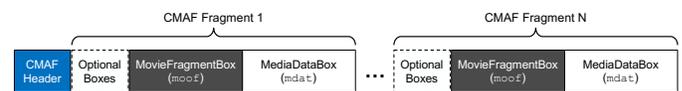

Fig. 11: Data structure of a CMAF track.

CMAF switching sets contain alternative CMAF tracks that can be switched and spliced at CMAF fragment boundaries for adaptive streaming. CMAF selection sets contain alternative CMAF switching sets that may include alternative content, *e.g.*, in different codecs, audio/subtitle languages and/or camera angles. Only one CMAF track in a CMAF selection set is intended for presentation at a time.

CMAF does not allow the use of multi-track (*i.e.*, multiplexed) ISO BMFF files with a shared presentation timeline. Instead, single-track files must be used. Thus, a streaming client must download each media component it selects (conforming to CMAF selection sets) separately and process each component as a separate CMAF track. CMAF provides the necessary hypothetical render model to align and synchronize these tracks. In this context, late binding refers to streaming clients being able to select and combine CMAF tracks in different combinations. For example, a video track can be used in combination with an English or Spanish audio track. If this video track had to be early bound with the English and Spanish audio tracks, one would have to create either a multiplexed file (containing the video plus the English and Spanish audio tracks) or two separate multiplexed files (one containing the video plus the English audio track and another containing the video plus the Spanish audio track). Even if we ignore the additional processing and packaging costs, the first approach leads to wasted downloads for the clients interested in listening to only one of the audio tracks, whereas the second approach requires duplicate storage on the servers and unnecessarily reduces the caching efficiency in the CDN.

*2) Chunked Encoding/Packaging:* As discussed in Section III-D1, live media is encoded and packaged into consecutive segments. However, as illustrated in Fig. 8 and 12, the latency, even under the best circumstances, cannot be less than one segment duration since the packager needs to wait for the entire segment data before it can complete packaging the segment and publish it. While one could choose to use shorter segments, decreasing segment duration becomes a limiting factor quickly as the encoding efficiency may drop substantially with shorter segments. The only other possible way to further reduce the latency is to perform encoding, packaging and delivery in shorter units.

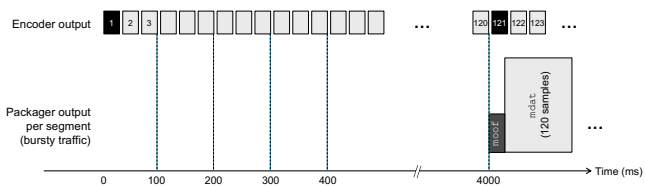

Fig. 12: Packaging of a non-chunked segment of four seconds. The video is encoded at 30 fps.

As the encoder generates an encoded frame (or a group of frames), it should be sent to the packager rather than waiting for the entire GoP to be encoded. This is called chunked encoding. As the packager receives the encoded frames, one or more frames are packaged into a CMAF chunk, and this is called chunked packaging. Fig. 13 illustrates the packaging of the four-second segment from Fig. 12 into 100-ms CMAF chunks (100 ms corresponds to three frames at 30 fps). As mentioned in the previous section, a CMAF chunk is immediately ready for delivery once it is packaged. This decouples the E2E latency from the CMAF segment duration and enables near-real-time delivery while the subsequent chunks of the CMAF segment are still being generated.

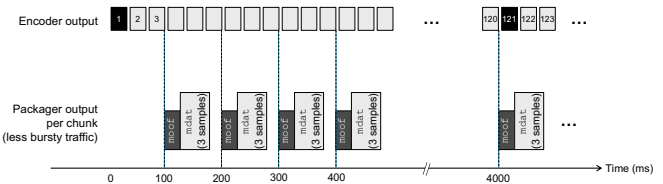

Fig. 13: Packaging of a segment of four seconds into CMAF chunks of 100 ms.

Chunk duration has an impact on the E2E latency. One may think that choosing the shortest possible chunk duration would result in the lowest latency. However, this may cause bandwidth measurement problems, as discussed in Section VII-A, and manifest refresh problems, as discussed in Section V-B2, that can eventually lead to higher latency or other performance issues. In the extreme case, each media sample could be packaged into a separate CMAF chunk, but this increases the packaging overhead, especially for the low-bitrate streams such as audio and subtitles.

*3) Chunked Delivery:* Chunking is made up of three sub-processes, chunked encoding, chunked packaging and chunked delivery, and these work closely to move live encoded media quickly through the LLL streaming workflow. Chunked delivery with HTTP is possible in different ways for different versions of the HTTP.

HTTP/1.1 employs a data streaming mechanism called chunked transfer encoding (CTE) (RFC 9112). CTE enables data of unknown size (dynamic data) to be transferred as a sequence of length-delimited chunks as they become available on the server. In LLL streaming, the media player sends a request for the live-edge segment that is still being encoded and packaged at the time of the request, and the server responds with whatever data it has so far while keeping the connection open until the segment is completed and fully transmitted. Once all the chunks of a segment are generated, they are combined to form the whole segment for storage and future requests. That is, if another media player sends a request for this already completed segment, chunked delivery is no longer needed, and the server's response includes the entire segment. This improves the network efficiency despite the added delivery latency.

Swaminathan *et al.* [196] were among the first to implement a CTE-based approach to decouple the live latency from the segment duration for LLL streaming. The approach analytically evaluated the latency in three live streaming methods, including (*i*) a segment-based method where the media player requested the segment only if it was fully available, (*ii*) a server-wait method where the media player requested a segment before it was ready at the server, and (*iii*) a chunked encoding method where the media player received chunks before the full segment was ready. Later, [128] and [75] investigated the adoption of CTE in live DASH delivery systems to reduce latency. The former introduced a new DASH timing parameter in the MPD that assists the player in its requests, namely the Availability Time Offset (ATO),
which specifies how much earlier the segments are available,

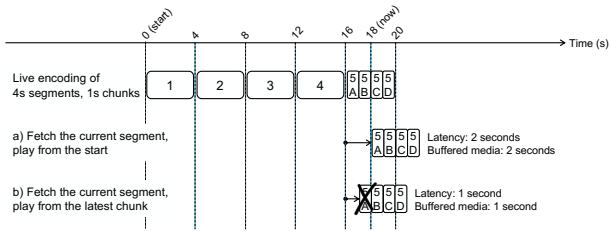

Fig. 14: Different playback strategies for chunked content, where each segment is four seconds and each chunk is one second long. Chunk fetching time is assumed to be negligible.

TABLE V: Field trial results for CMAF-based LLL streaming solutions. These data came from field trials conducted by Harmonic and Akamai. They tested an end-to-end CMAF-based LLL streaming setup using two deployment methods (on-premise and cloud) and multiple network conditions (wired and mobile LTE). Latency measurements were recorded using 100-ms chunks. The results demonstrate the effectiveness of CMAF for achieving low latency.

|  | **On-premises** | **Public Cloud** |
| --- | --- | --- |
| Network | Wired (Wireless) | Wired (Wireless) |
| CDN | Akamai | Akamai |
| Player | Fixed (Mobile) | Fixed (Mobile) |
| E2E Latency | 5.5 seconds (7.5 seconds) | 7 seconds (9.5 seconds) |

while the latter developed an HTML5 player with CMAF-enabled LLL support. Shuai *et al.* [183] also analyzed the main contributors to the live latency and presented a mathematical model that approximated the buffering delay's lower bound to reduce latency.

It is important to note that for the chunked delivery to be effective, the origin server and any other intermediate cache within the CDN must support and enable CTE, as shown in Fig. 15. As chunks are received on an intermediary, they can be forwarded downstream to minimize the latency. If any intermediary waits until the segment is fully received, the latency increases by the segment duration. On the other hand, H2 (RFC 9113) and H3 (RFC 9114) do not support the CTE method as they already use a streaming transport, so chunked delivery is inherent to them.

Fig. 14 illustrates a live streaming scenario with the chunked content (four-second segments, one-second chunks) and CTE enabled. A media player joins the live session 18 seconds into the content. There are two common strategies the media player can adopt:

- The media player determines the live-edge segment (*i.e.*, segment #5) and requests it. It can commence the playback as soon as the first chunk (chunk #5A) is received, achieving a latency of two seconds.
- If the media player is more adventurous, it requests segment #5 and decodes chunks #5A and #5B but starts the playback from chunk #5B, reducing the latency to one second.

To confirm the efficiency of CMAF-based LLL streaming solutions in assuring low latency, Harmonic and Akamai have conducted a variety of field tests [97]. These tests covered two deployment methods (on-premise and cloud) and various access network conditions (wired and mobile LTE). Using two-second segments and 100-ms chunks, the average E2E latency was 5.5 seconds, similar to what is typically achieved in linear broadcast. The results are summarized in Table V and conclude that on-premise deployment on wired networks can provide a low latency approaching that of broadcast.

### B. Protocol Extensions

*1) Low-Latency Extensions for DASH:* Low-latency DASH (LL-DASH) was introduced in 2017 through interoperability (IOP) profiles and specification extensions. In early 2020, the DASH Industry Forum (DASH-IF) published a new change request to the DASH IOP guidelines on low-latency modes for DASH [67]. This extended standard for DASH, based on the earlier joint work between the DASH-IF and Digital Video Broadcast (DVB) (available in [70]), presents features like Resync elements for fast stream joining, implementation details and recommended settings for multi-bitrate encoders, origin servers, CDNs and media players.

LL-DASH achieves low latency thanks to two essential technologies: chunked encoding/packaging (through CMAF) and chunked delivery. First, CMAF alone does not reduce latency, but it makes LLL streaming possible by splitting media segments into smaller, non-overlapping chunks. Second, chunked delivery helps deliver chunks to the media players as soon as they are ready without having to wait for the entire segment to be completely packaged. In LL-DASH, the MPD of a live session indicates when a segment will start becoming available. Therefore, the LL-DASH client can request the live-edge segment immediately. This feature requires accurate client-server clock synchronization (using the `UTCTiming` element and Network Time Protocol (NTP; RFC 5905)).

The `ProducerReferenceTime` element in the MPD provides a correlation between media timestamps and wall-clock production time. First, this helps LL-DASH clients avoid possible clock drifts. Second, it enables measuring and controlling the latency (as optionally specified using the `ServiceDescription` element indicating the minimum and maximum latency and playback speed values acceptable for the service). The same correlation information can also be provided as part of the media segments in the `prft` box, as defined in [108].

It is worth mentioning that the encryption process using DRM might introduce extra delays that may impact the latency as the media player must first acquire a license for the playback to start. In LLL streaming, the target buffer is small, and making additional requests to exchange the encryption keys could cause a delay of a few hundred milliseconds (or sometimes more), impacting both startup delay and latency. For LL-DASH, it is recommended to have DRM initialization information in the MPD. This technique allows the media player to start negotiating for a license before any media segments are to be loaded.

*2) Low-Latency Extensions for HLS:* Apple released its extension of the HLS protocol for low latency, termed LL-HLS, in 2019. The specification [174] was updated in April 2020 and aimed to provide the same scalability as HLS, with a latency of two to six seconds as compared to 12 to 30 seconds achieved by the traditional HLS. When Apple announced LL-HLS, the industry was surprised by Apple's decision not to use CTE but to use H2 (RFC 9113) with server push capabilities along with partial segments (also called parts). This meant LL-HLS introduced significant new requirements for the origin servers and CDNs.

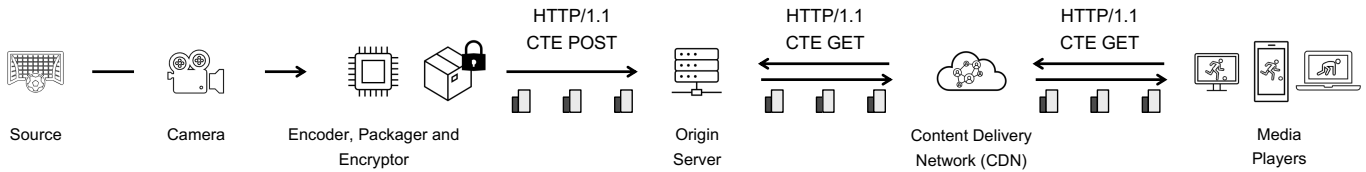

Fig. 15: CTE delivery.

LL-HLS uses the traditional playlist approach, meaning that the playlists have to be updated more frequently, and the clients have to request every partial segment separately to achieve low latency. Depending on the choice of partial segment duration, the request rate of an LL-HLS client can be substantially higher. Live events draw a big audience; there will be many concurrent clients, each with a high request rate, and this can be a major concern for the CDN and network service providers due to increased network traffic.

On the more positive side, LL-HLS playlists describe the internal structure of the segments. While LL-HLS does not require the partial segments to start with a key frame, using the `INDEPENDENT=YES` attribute, ones containing a key frame are indicated in the playlist. This way, the LL-HLS client can determine the best partial segment to fetch upon starting up or switching to a different variant stream. LL-HLS clients also experience more deterministic startup delays. The `PART-HOLD-BACK=<seconds>` attribute in the playlist tells an LL-HLS client to start up that many seconds behind the live edge, unless the value specified by `AVPlayerItem.configuredTimeOffsetFromLive` is greater. The recommended value for `PART-HOLD-BACK` is at least three times the maximum partial segment duration.

In essence, the new features and requirements introduced by the low-latency extensions for HLS can be summarized as below:

- A media segment is divided into partial segments (*e.g.*, with 200 milliseconds each) that conform with the CMAF format. Each part is listed separately in the playlist, and parts do not have to have identical durations. In contrast with segments, listed parts can disappear from the playlist while the segments containing the same data remain available for a longer time.
- LL-HLS clients initially request the master playlist and the media playlists and then request only a delta update as parts are added. A delta update contains only the last few segments, along with the parts at the tail of the playlist. Delta updates are smaller in size, and hence, reduce the amount of traffic for playlist refreshes.
- When an LL-HLS client issues a GET request for a playlist update, it can add special query parameters, called delivery directives, to specify that it wants the playlist response to include a future segment. In this case, the server holds onto (blocks) the request until a version of the playlist that contains that segment is available. This feature, called blocking playlist reloads, helps eliminate constant polling of the servers. Coalescing duplicate blocking requests also helps reduce the load on the CDNs. For some CDNs, the impact of a large amount of blocking connections may still be a concern, as they are generally designed to serve requests as quickly as possible.
- A playlist may include preload hints, allowing the LL-HLS clients to anticipate what data must be downloaded (using the EXT-X-PRELOAD-HINT tag), which reduces overhead and unnecessary RTTs. For example, an LL-HLS client can issue a GET request for a hinted media part in advance, and the server responds to the request as soon as that media part becomes available.
- In LL-HLS, each media playlist contains information about the segments in other renditions listed in the master playlist to allow faster switching with the minimum number of round trips. The rendition reports are carried in the EXT-X-RENDITION-REPORT tag and provide information such as the last media sequence number and part currently in the media playlist of that rendition.
- Servers can provide unique identifiers for every version of a playlist, easing the cacheability of the playlist and avoiding stale information in the caches.

LL-HLS provides backward compatibility where the traditional HLS clients can play the same LL-HLS streams, albeit at a higher latency. Therefore, such a feature allows the reuse of the same stream for both LLL and non-LLL streaming, regardless of the platform and availability of an HLS client with low-latency support.

To summarize the discussion, Table VI lists the different characteristics of the low-latency variants of DASH and HLS. The evolution of LL-HLS and an early 1:1 performance comparison between LL-HLS and LL-DASH have also been presented in [73].

*3) Serving The Same Content with LL-DASH and LL-HLS:* While the introduction of LL-DASH and LL-HLS allows for reduced latency compared to their legacy counterparts, the problem of inefficiency in alternatively packaged content remains. Both LL-DASH and LL-HLS were developed independently at their inception. As seen in the previous sections, some differences exist in their modes of operation and implementation details, which may result in content delivery systems having to deploy separate streams to support these formats. In this regard, Law *et al.* [126] proposed a solution that achieves interoperability among the four types of streaming clients (legacy DASH, legacy HLS, LL-DASH and LL-HLS). In essence, the solution combines the CMAF format and chunked encoding/packaging with byte-range addressing to satisfy each standard's requirements. Such a solution not only improves cache efficiency by allowing a single set of media objects to be distributed in multiple formats but also achieves other performance gains, as further elaborated below.

TABLE VI: Comparison of low-latency extensions for HAS. Filled and empty circles refer to present and absent characteristics, respectively. Here, LHLS represents the community version of the low-latency HLS, which was introduced by Twitter's Periscope in 2016 with similar features as LL-DASH [166] and later improved by the streaming community in 2018. LHLS is no longer used but is included here for completeness.

|  | LL-DASH | LHLS | LL-HLS |
|---|---|---|---|
| Chunked encoding/packaging | ● | ● | ● |
| Use of CTE | ● | ● | ○ |
| Defined segment structure | ◐ | ○ | ● |
| Per-chunk manifest update | ○ | ○ | ● |
| Line-speed delivery | ○ | ○ | ● |
| Deterministic startup | ○ | ○ | ● |
| Smart origin to modify manifest | ○ | ○ | ● |

LL-DASH and LL-HLS use CMAF to generate partial segments, which are referred to as chunks and parts, respectively. Each partial segment can be addressed discretely via a unique URL or optionally as a referenced byte range into a media segment. While most HTTP-based LLL streaming solutions have focused on the discrete addressing mode, byte-range addressing brings several performance benefits:

- The CDNs can serve DASH and HLS clients with or without the low-latency mode using the same stream, which improves the origin storage by a factor of three and the CDN's cache efficiency.
- The request rate for LL-HLS clients can be reduced depending on the part and segment durations. For example, the LL-HLS, by design, generates one request per part for each media type, which creates a significant overhead for the CDN. Having byte-range addressing, an LL-HLS client needs only to make one request per segment for each media type, improving CDN scalability and reducing the overall system cost.
- Early versions of LL-DASH or LL-HLS encoders produced partial segments that were all independent, *i.e.*, each one had a key frame, and contiguous segments with a single key frame for legacy DASH and HLS clients. This increased the CDN cache storage and broke the portability of storing a single object in the cache from which one could serve both partial and complete segments. In order to achieve a unified cache, the segment structure must be a direct concatenation of its partial segments. Such a technique halves the cache footprint and far outweighs the small encoding efficiency gains achieved by having two bit-different objects (partial segments and the complete segment).
- An LL-HLS client under steady-state playback does not need to make any byte-range requests against the origin server, even when range-based addressing is used in the playlist. This removes the Cross-origin Resource Sharing (CORS) preflight requirements for browser-based clients[3], improving the latency with which playlists and segments can be returned.

Interested readers may refer to [10], [126] for details on the design, implementation and field evaluation of such a cross-format HTTP-based LLL streaming system.

*4) High Efficiency Streaming Protocol (HESP):* HESP has been designed to maintain sub-second latency, compared to legacy DASH and HLS [103], [191], [217]. HESP has been shown to achieve an E2E latency of as low as a few hundred milliseconds and short startup delays (*i.e.*, fast channel changes and short seeking times). In contrast to the segment-based approaches, HESP employs a frame-based streaming approach, enabling a better trade-off between the live latency and startup delay. HESP uses two distinct types of streams, as shown in Fig. 16:

- Initialization Streams: These contain only key frames. Initialization streams remain unused until a new stream is initiated.
- Continuation Streams: These are regularly encoded streams, ensuring playback continuity after any initialization stream event.

The corresponding media in both streams must be issued with synchronized presentation timestamps. Using CTE in conjunction with byte-range requests, HESP enables rapid stream initiation and bitrate adaptation to changing network conditions, as shown in Fig. 16. This reduces media player buffer requirements for an equivalent viewer QoE, leading to lower latencies and faster channel changes. There are several implementation details about the encoding and packaging of the initialization and continuation streams. More details can be found in the Internet-Draft [163].

*5) Low-Latency Low-Delay Extensions for DASH (L3D-DASH):* This DASH profile is currently under development in MPEG and will soon be published in the sixth edition of the DASH standard. This profile allows fast switching and fast tune-in, both of which improve the viewer experience and enable fast content switching (*e.g.*, for unconditioned ad insertion). Furthermore, this profile allows for the use of variable-duration segments, and thus, improves the compression efficiency of the media segments.

## VI. GOING BEYOND HTTP TO FURTHER REDUCE LATENCY

In this section, we present a brief overview of the developments in using real-time communication protocols together with DASH to improve interactivity in LLL streaming. We also summarize the preliminary discussions that led to the forming of the new MOQ working group in the IETF.

---
[3]Available [Online]: https://developer.mozilla.org/en-US/docs/Web/HTTP/CORS

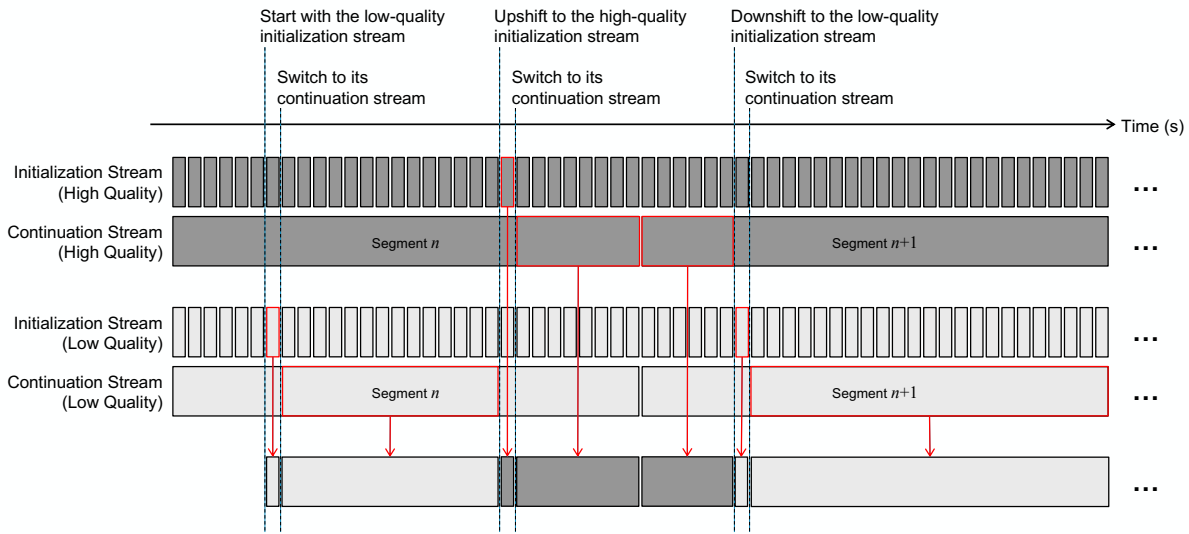

Fig. 16: Illustration of a HESP media player quickly starting the playback with low latency, then upshifting to the high-quality stream in the middle of segment $n$ and downshifting to the low-quality stream in the middle of segment $n + 1$.

### A. WebRTC and DASH

WebRTC [25], [215] is a set of W3C and IETF standards that delivers real-time content to viewers, typically with an end-to-end latency of under half a second. It builds upon the prior RTP protocol suite (see Section III-B). Support for WebRTC is built into all modern Web browsers across desktop and mobile devices, and the library functions allow for video, audio and data streaming. While the original focus of WebRTC was videoconferencing, it is increasingly being used today for real-time streaming of premium content because of its ultra-low latency features. These features may enable several new user experiences, especially those involving user interactivity that is not easy to deliver or even possible with traditional broadcast or HTTP-based streaming delivery protocols alone.

As interactive premium content services scale up, the integration of WebRTC with DASH is deemed essential. However, initial attempts at this highlighted some gaps in WebRTC and DASH. In May 2022, the DASH-IF invited experts to study the details of a WebRTC-based premium streaming ecosystem. The goals were to recognize synergies between WebRTC and DASH, explore opportunities and identify open issues. This study resulted in an initial report [69] that outlined example use cases, developed key performance indicators and provided a baseline architecture. The report also outlined the technical work needed to integrate WebRTC with DASH. An initial proof-of-concept implementation based on the report's findings is available in [25].

### B. Media over QUIC (MOQ)

The IETF formed a new working group called Media over QUIC (MOQ) in 2022 to explore new opportunities for low-latency streaming using QUIC [104]. Its charter is to develop a scalable and efficient low-latency media delivery solution for media ingest and distribution in browser and non-browser environments. Targeted applications include live streaming, cloud gaming, remote desktop, videoconferencing and eSports. The work is still in its infancy, but media over QUIC running in an H3 or WebTransport [214] environment could be a game-changer [111].

MOQ defines a latency-configurable delivery protocol for sending media from one or more producer(s) to consumer(s) through relays using either WebTransport (in browsers) or raw QUIC (elsewhere). Relays take incoming media and forward it to one or more relays or consumers to scale media distribution without needing a separate encoding for each consumer. They decide what to send in what order or discard based on specific metadata exposed in the envelope of the incoming packets to respond to congestion and satisfy the application's latency requirements. Consumers can also trade off quality with latency by choosing how long to wait for media based on their network conditions and desired user experience. Fig. 17 shows that MOQ is designed to be a standard media protocol stack, which can facilitate two primary functions: (*i*) enabling live streaming of events, news and sports with enhanced interactive capabilities, and (*ii*) scaling up real-time media applications to cater to larger audiences.

The MOQ proposals are still under discussion and the protocol details are still evolving. Some initial results and benchmarks have been reported in [89]–[93].

## VII. MEDIA PLAYER ENHANCEMENTS AND EXAMPLES

Legacy media players have to be re-designed to support LLL streaming. Among the components and functionalities of a media player, we explain the new ones and the ones needing modifications in this section. This section also provides an evaluation of the existing open-source media players.

### A. Bandwidth Measurement

In traditional non-LLL streaming, measuring bandwidth is straightforward. Consider Fig. 18 where the media player fetches the segments already packaged and available on the server. Therefore, the fundamental equation of bandwidth measurement ($Q_i/\tau_i$) provides a good approximation for the available bandwidth, where $Q_i$ denotes the size of segment $i$ and $\tau_i$ denotes the download time for segment $i$ (*i.e.*, the time difference between when the request was sent and when the segment was fully received).

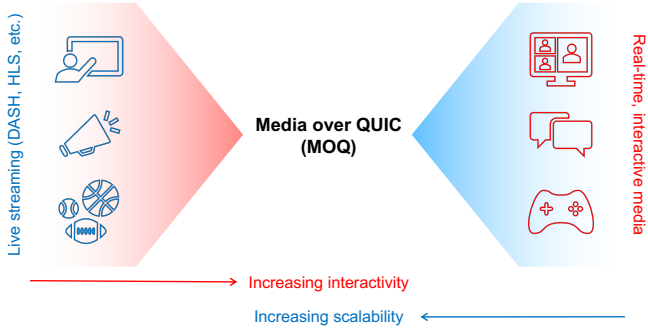

Fig. 17: MOQ is envisioned to improve both scalability and interactivity in multimedia applications. Adopted from [89].

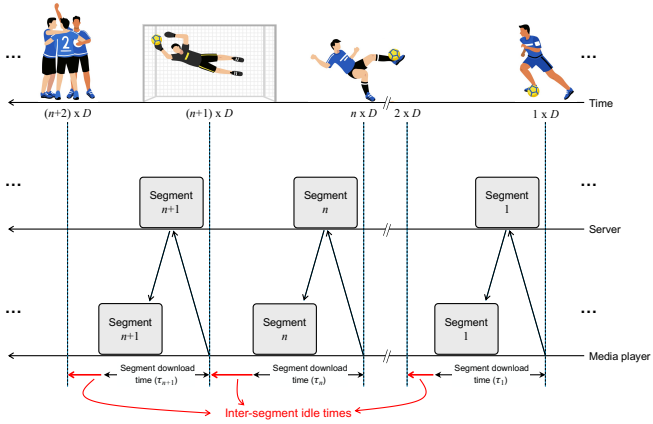

Fig. 18: The $Q_i/\tau_i$ formula works well in non-LLL streaming. $Q$, $D$ and $\tau$ indicate the segment size, duration and download time, respectively. Segments are colored differently. Not to scale, for illustration purposes only.

In contrast, when using chunked encoding/packaging and delivery for LLL streaming as depicted in Fig. 19, the media player fetches the live-edge segment where the chunks are transmitted in two different transmission phases [41], [42], [133]. First, in the *network-limited* phase, the available chunks of the requested segment are sent by the server (*e.g.*, first the origin and then the CDN server) at the full network speed to the media player. Then, in the *source-limited* phase, the remaining chunks of the requested segment are sent to the media player as they become available. Assuming the network is fast enough, idle time is introduced between the chunks as they are sent from the server. This way, the chunks are sent with relatively more consistent timing ($\tau_i \sim D$). If the network is not fast enough, then the transmission is network-limited, and the ABR scheme naturally needs to downshift to a lower bitrate. In either case, the total segment download time, from the first chunk to the last chunk (including idle times in between), cannot be significantly shorter than its duration. Thus, a media player naively using the $Q_i/\tau_i$ formula will compute a value nearly equal to the segment encoding bitrate. This prevents the media player from switching to a higher bitrate level, even if the network has enough bandwidth to support a higher level.

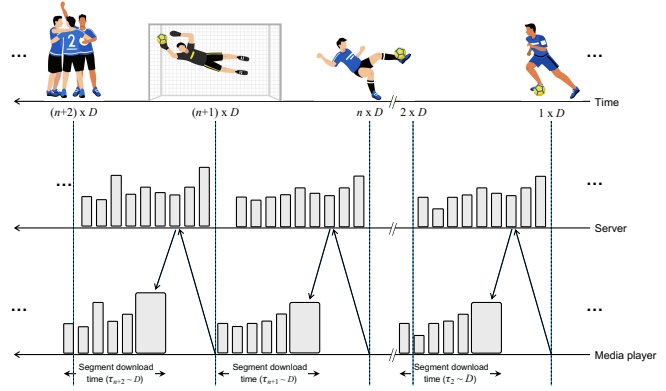

Fig. 19: The $Q_i/\tau_i$ formula does not work well in LLL streaming. Segments (colored differently) consist of bars, each indicating a chunk, with its size being proportional to the bar's area. Not to scale, for illustration purposes only.

To understand this better, refer to Fig. 20. After segment $i$ is fully downloaded, the media player uses the $Q_i/\tau_i$ formula to measure the bandwidth, where

$$Q_i = \sum_{n=1}^{z} q_i^n \text{ and } \tau_i = e_i^z - b_i^1. \quad (1)$$

Here, $q_i^n$ is the size of the $n^{th}$ chunk, $z$ is the number of chunks in segment $i$, and $b_i^n$ and $e_i^n$ are the download beginning and end times of the $n^{th}$ chunk, respectively. In (1), the segment download time ($\tau_i$) includes the inter-chunk idle periods, potentially resulting in incorrect bandwidth measurements.

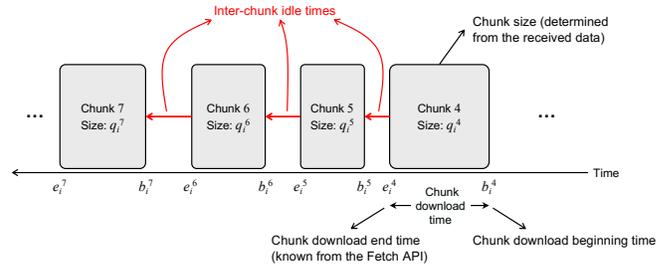

Fig. 20: The challenge of bandwidth measurement in LLL streaming with chunked encoding/packaging and delivery.

A new chunk-aware bandwidth measurement technique taking the inter-chunk idle periods into account is clearly needed. The first solution in this space was [41], which proposed the ABR for Chunked Transfer Encoding (ACTE) method. ACTE used a per-chunk sliding window moving average (SWMA) bandwidth measurement algorithm and predicted the future bandwidth based on a recursive least squares (RLS) algorithm. For each segment download, measuring the bandwidth involved two steps: (*i*) calculating the download rate of each chunk, which equaled its size divided by this chunk's end time minus the previous chunk's end time, while (*ii*) filtering out the chunks with a non-negligible idle period (the chunks whose download rate was within ±20% of the average segment download rate).

ACTE achieved good performance in case of short inter-chunk idle times. However, it suffered bandwidth underestimation problems when the idle times were longer. To rectify this issue, first, the Low-on-Latency (LoL) solution [133] and then its extension, LoL$^+$ [32], were developed. The LoL and LoL$^+$ solutions are learning-based and work efficiently for both short and long inter-chunk idle times, thanks to:

- Chunk Boundary Identification: The module first successfully identifies each chunk's beginning and end times by capturing the `moof` box in the downloaded data. It leverages the fact that the chunks are transmitted as separate HTTP chunks through CTE. Via the Fetch API, this module uses `Streaming Response Body`[4], which allows tracking the progress of the chunk downloads and parsing the chunk payloads in real time. When a `moof` box is captured, the module stores the time as the beginning time of the chunk download. Then, it stores the end time of the chunk and its size (in bytes) when the chunk is fully downloaded. The inter-chunk idle times can be removed entirely from the segment download time with the accurate beginning and end times for each downloaded chunk.
- Chunk Filtering: Upon receiving all the chunks of a segment, the filtering process is triggered to remove noisy measurements. In this case, the first and the last chunks are ignored in the bandwidth calculation to omit any transient outliers.
- Segment Bandwidth Smoothing: After the filtering process is completed, this module first computes the segment download time ($\tau$) using only the chunks that passed the filtering process and then uses the SWMA-based smoothing algorithm to compute the measure of the bandwidth.

Later, [33] found that the LoL and LoL$^+$ bandwidth measurement modules could sometimes perform poorly if the network conditions showed significant variations. To address such cases, the authors proposed the Automated Model for Prediction (AMP) that encompassed techniques for bandwidth prediction and model auto-selection designed explicitly for LLL streaming with CTE. The results from the AMP study showed that using several prediction models and automatically selecting a good one during the streaming session was beneficial. Different prediction models have their own characteristics and strengths, and hence, their effectiveness varies under different network conditions and user mobility types.

[4]Available [Online]: https://developer.mozilla.org/en-US/docs/Web/API/Streams_API/Using_readable_streams

We note that since LL-HLS uses a separate request for each partial segment, the presented bandwidth measurement problem does not apply to LL-HLS. We also note that other bandwidth measurement methods exist, including the traditional active probing methods such as packet pair/train dispersion probing that is often used to estimate the end-to-end capacity of a network path [21], [172], and Will's simple side-load (WSSL) method, which retrieves a fully available object (such as a previous segment) along with the live-edge segment for bandwidth measurement purposes [41]. However, active probing methods may be less desirable in media streaming applications as these methods introduce additional traffic to the network (which already is a scarce resource in streaming) [50]. They are often not well-adapted to handle streaming traffic's periodic and bursty nature, either [212]. To mitigate these issues, [131] designed an active probing algorithm that probes using incremental data rate (instead of piggyback traffic) that backs off when network congestion is detected.

### B. Rate Adaptation

Traditional (non-LLL) ABR schemes [40] cannot sustain smooth playback with tiny playback buffers as they do not provide enough protection against sudden bandwidth drops. To be able to deploy LLL streaming services at scale, several considerations should be taken into account:

- Small Buffer Levels: The duration of the media in the playback buffer should not be larger than the target latency. That means when the target latency is in the one-second range, media players have just a few hundred milliseconds to react to bandwidth variations. Since real-world network conditions can vary significantly, especially on mobile networks, media players are challenged with balancing the number of bitrate switches against preventing rebuffering.
- Variable Playback Speeds: Changing playback speed during the live streaming session can be helpful to maintain the target latency. When the media player falls significantly behind the live edge (*i.e.*, the instantaneous latency increases), it can accelerate the playback to catch up and prevent its buffer from growing arbitrarily as long as media arrives fast enough. However, in such cases, buffer-based ABR schemes like BOLA [192] may struggle since the buffer level may not grow large enough for upshifting.
- Fairness or Controlled Unfairness: Media players watching the same stream (or a different stream from the same server) or the ones sharing network links will be competing for the available bandwidth unless the supply of the server and network capacity is guaranteed to be more than the demand. Previously, [12] showed that one or more problems, such as bitrate oscillations, network resource underutilization, sustained starvation and (bitrate) unfairness among the media players, were inevitable. If a media player is too greedy or behaves incorrectly, it risks degrading the experience of other players. On the other hand, not all media players run on devices with identical capabilities or display sizes, nor do they consume the same or the same type of content (*e.g.*, sports, news or a movie). Thus, fair-share bandwidth allocation is largely a bad idea [34]. One rather needs controlled unfairness instead of absolute bitrate fairness not only to avoid bitrate oscillations but also to achieve quality fairness among the media players [26], [27].
- Playback Synchronization: Many applications in live streaming rely on synchronized playback, such as synchronizing across multiple users co-watching the stream to maintain fairness and trust (and meet legal obligations in some cases) in betting applications and synchronizing across multiple camera angles of the same content that a user is consuming concurrently.

The recent ABR schemes designed explicitly for LLL streaming can be categorized as heuristic-based or learning-based. Heuristic-based schemes (*e.g.*, LoL heuristic-based [133], Stallion [94] and Fleet [130]) use mathematical rules or approximations to determine the next best bitrate to request from the server. In an earlier study, Peng *et al.* [165] used the hybrid control theory to formulate the ABR decision problem for low latency. It implemented three essential rules: a heuristic-based playback rate control, latency-constrained bitrate control and QoE-oriented adaptive frame dropping. It also leveraged Kaufman's adaptive moving average (KAMA) algorithm to predict the bitrate for the next segment for better rate adaptation.

In contrast, learning-based schemes (*e.g.*, LoL learning-based [133], L2A [117], Fugu [227] and ALVS [162]) learn from changes in the system environment to make decisions based on past data. More specifically, leveraging advances in deep reinforcement learning (DRL), Jiang *et al.* [114], Hong *et al.* [99], Zhao *et al.* [236], and Wang *et al.* [218] proposed the following ABR schemes for LLL streaming: HD3, CBLC, L2AC and BitLat, respectively. HD3 [114] is a DRL algorithm that combines discrete actions (bitrate and target buffer level) with continuous actions (latency limit values) to optimize video quality while minimizing E2E latency. CBLC [99] uses a deterministic policy gradient for continuous ABR decisions and latency control tasks. L2AC [236] trains several separate DRL-based models on various network conditions with a reward function that maximizes QoE and penalizes latency. Then, it uses the model ensemble technique to combine these models to improve the overall performance. BitLat [218] develops a latency-aware RL-based solution to select a suitable bitrate while maintaining low latency. Further, it includes a dynamic reward method in the model to boost performance.

Each of these ABR schemes has its own merits in that it could achieve better performance in some scenarios or under certain network conditions but not in others. This suggests that having an ABR switching module that implements multiple ABR schemes and could select the best one based on the current requirements and conditions may help to leverage these schemes better [133]. In selecting a suitable ABR scheme for the job, one should consider the following aspects:

- Predictability of the Media Characteristics and Network Conditions: Bitrate selection depends on how much certainty we have about the content being streamed and the conditions of the network to which the media player is connected. If chunk/segment sizes vary significantly over time and are not known a priori (within an error margin) or if the available bandwidth varies too much, and hence, becomes difficult to predict, then a learning-based scheme is a better choice as heuristic-based schemes may perform poorly with larger prediction errors. This arises from the inherent limitation of heuristic-based schemes, as they cannot consistently adapt to varying network conditions.
- Significance of the QoE Model: In systems where the desired QoE model is not well-defined, a learning-based scheme is a better choice as it can individually learn from each QoE metric and make nuanced trade-offs. This contrasts with heuristic-based schemes, which rely heavily on the QoE model function.
- Performance Comparison of the ABR Schemes: At times, the choice and performance of an ABR scheme may not be clear. Conducting extensive experiments to evaluate the performance of each ABR scheme under different conditions would be the ideal way of understanding their performances and determining which one to use. This approach is tedious but also needed as we explore the idea of ABR switching, as explained next.
- Switching between the ABR Schemes: This is achieved through a data-driven AI-based switcher, which takes many inputs from the environment (*e.g.*, network conditions, player status, stream complexity and LLL requirements) and auto-selects the best-performing ABR at run-time, and keeps switching between these ABR schemes during the streaming session for better performance.

*C. Playback Speed Control*

Using the conventional playback speed control techniques without adapting them for LLL streaming or not using any such techniques may cause two major issues: more frequent rebuffering events (as the buffer sizes are smaller) and an uncontrolled increase in latency (as rebuffering durations keep adding up to the latency). The main goal of the playback speed control is to determine a rendering speed that ensures a latency close to the target while taking a calculated risk of rebuffering [14], [66], [165].

[32] developed a hybrid playback speed controller based on two variables: (*i*) the difference between the current and target latency, and (*ii*) the difference between the current playback buffer level and a pre-specified/configured safe threshold (all measured in units of time). Specifically, the media player considers the following cases when dynamically adjusting the playback speed:

**Case 1:** The current buffer level is below the safe threshold. In this case, the playback should be slowed down.
**Case 2:** The current buffer level is equal to or above the safe threshold and further:
   **2a:** The current latency is close enough to the target latency, then the media is played at the normal speed (1x).
   **2b:** The current latency is lower than the target latency, then the media is played at a slower speed.

TABLE VII: Example playback speed adjustment for three different cases in LLL streaming [32], [97]. This is a theoretical example to illustrate the logic behind a playback speed control system. Specifically, it demonstrates how a media player might adjust speed depending on the comparison of buffer occupancy and current and target latencies. The example validates the discussion that a constant playback speed is not sufficient for LLL streaming and the playback speed has to be changed appropriately.

|  | Case 1 | Case 2a | Case 2b | Case 2c |
|---|---|---|---|---|
| Target Latency | | 1.5 seconds | | |
| Minimum Buffer Threshold | | 0.5 seconds | | |
| Allowed Playback Speed | | 0.5-1.5x | | |
| Current Buffer Level (seconds) | 0.3 | 0.5 | 0.7 | 0.7 |
| Current Latency (seconds) | Irrelevant | 1.45 | 1.3 | 2.0 |
| Playback Speed (x) | 0.75 | 1.0 | 0.85 | 1.1 |

**2c:** The current latency is higher than the target latency, then the media is played at a faster speed.

Here, the playback speed is determined from a configurable speed range (*e.g.*, 0.5 − 1.5×) based on the deviation between the current and target latencies and between the current buffer level and safe threshold. However, to reduce the impact of playback speed changes on the QoE, the controller should try to stay at the normal speed as long as possible. Table VII illustrates an example of the above-mentioned cases. Other works that explored the use of playback speed adaption to manage latency include [130], [233], [234].

Varying the playback speed must still consider the synchronization between the media streams sharing the same presentation timeline. For the video, the process is straightforward since frames can readily be displayed, either faster or slower than normal. For the audio, time-scale modification [160] can be applied to enable slower or faster playback while preserving the pitch of the signal. Note that all the media is still played out (no skips) despite the playback speed changes.

One problem with playback speed adaptation is that it may introduce discernible artifacts to the media, which could sometimes annoy the viewer (*i.e.*, the viewer can tell that the video/audio has been sped up or slowed down). More sophisticated playback speed control solutions have been developed to mitigate this, such as the content-aware playback speed control (CAPSC) algorithms [11], [13] that intelligently vary the playback speed based on the content to be played at that instant such that the artifacts introduced are less noticeable (or almost unnoticeable) by the viewers.

### D. Playback Buffer Management

LLL streaming requires robust buffer management that can tolerate sudden and significant changes in network conditions. Anytime, the playback buffer level should be above the given safe threshold. If one is provided, the buffer level should also be below the maximum latency acceptable for the service. This can be achieved by configuring the media player settings properly. The default settings for most media players are set to favor smoother playback with less rebuffering, meaning that they prioritize having more media data in the playback buffer over operating at a lower latency. In this section, we describe some guidelines for attuning the media player settings for LLL streaming:

- Buffer Sizing and Initial Playhead Positioning: Media players, by default, are configured to download a specific number of segments (assuming a particular minimum segment duration of several seconds) and/or a specific number of seconds of media before the playback can commence. For example, DASH-based media players may be informed to download a specific duration of media through the `minBufferTime` attribute in the manifest [109].

  Recall that the startup delay is distinct from the latency (refer to Section I-C). The media player can start downloading from earlier segments to fill the necessary buffers and start the playback sooner. However, in this case, the latency will not be necessarily low. Alternatively, the media player can download a segment closer to the live edge to reduce the latency, but the startup delay may be larger this time. Therefore, one must coherently configure the media player's settings regarding the startup delay, buffer sizing and latency and make sure that one setting does not override another.

- Sticking to the Live Edge: When a media player stalls for a certain amount of time and then resumes the playback from where it left off, the instantaneous latency will at least increase by the rebuffering duration. By controlling the playback speed, the media player can attempt to close the gap to the live edge, provided the gap is not significant to start with. If it is, the media player can instead jump forward in the presentation timeline (*i.e.*, by skipping media) to reduce the latency back to the target value. This behavior has been available in the dash.js reference player since v2.9.2.

- Resilience to Late/Missing Segments: When a media player receives an HTTP 404 error for a requested segment, it can be due to three reasons. First, the media player might have sent the request earlier than it should have. Second, the CDN and the origin server may not have started receiving the requested segment because of an unexpected delay in the media contribution/ingest stages (refer to Fig. 3). Third, the segment might have been corrupted or lost in the upstream. Upon receiving the 404 error, the media player can wait for a bit and retry the request. Alternatively, it can request the corresponding segment from a different representation. If none of the temporally same segments is available, the media player must skip this segment and continue with the subsequent segment(s).

  If a media player occasionally or consistently keeps sending early requests, this causes a major problem in media distribution. Even after the particular segment becomes available on the origin server, the 404 error will be cached by downstream CDN servers, and all the media players requesting this segment will also receive the 404 error until the cached error message expires. In other words, a misbehaving or faulty media player could introduce a problem to all the other media players watching the same LLL stream. To avoid this *avalanche* effect, some CDNs turn off caching the 404 errors at the risk of increasing the request traffic back to the origin server. Another (but costly) solution is to implement the blocking request feature in the origin server, which is already required to support LL-HLS. On the LL-DASH side, the media players must be time-synchronized with the packagers that prepare the media segments and manifests. An NTP server (*e.g.*, https://time.akamai.com/) can be signaled in the manifest (using the UTCTiming element), and this way, clock drifts can be avoided and the media players can be kept tightly synchronized.

For both LL-DASH and LL-HLS, enabling truly fault-tolerant media contribution and media ingest stages is critical. This can be achieved by using multiple encoders, packagers and origin servers in the workflow so that multiple interchangeable and synchronized outputs can be produced. For details on the ongoing standardization, see [145], [154].

### E. Examples of Open-Source Media Players

There are multiple open-source media players available. These players have support for DASH, HLS or both. Among the popular open-source media players, we mention:

- The dash.js reference player [66] is an initiative of the DASH-IF to establish a production quality framework for building video and audio players that playback DASH content using client-side JavaScript libraries leveraging the Media Source Extensions (MSE) and Encrypted Media Extensions (EME) APIs defined by the W3C.
- Shaka player [84] is a JavaScript library for HAS, developed by Google. It plays adaptive media formats (such as DASH and HLS) in modern browsers using the MSE and EME APIs.
- hls.js player [1] is another JavaScript library developed by multiple companies. It relies on HTML5 media and MSE for playback. It works by transmuxing MPEG-2 Transport Stream (TS) and AAC/MP3 streams into ISO BMFF (MP4) fragments.
- ExoPlayer [83] is a DASH, HLS and Microsoft Smooth Streaming (MSS) media player for Android, developed by Google. It provides an alternative to Android's MediaPlayer API for playing audio and video locally and over the Internet.

Over the last few years, these media players have been extended to support LLL streaming protocols as described in Table VIII. Following the AWS recommendations [14], we conducted experiments over the AWS Elemental Media Services infrastructure [15] to investigate the achieved performance for these open-source media players with low-latency mode enabled (CMAF+CTE), *i.e.*, dash.js (v3.2) [66], Shaka (v3.1) [84], hls.js (v1.0.0-rc.4) [1] and ExoPlayer (r2.13.2) [83], using four main scenarios. The main goals of these experiments are to show the following: (*i*) the impact of media player parameters on latency and quality, (*ii*) the performance of the E2E workflow architecture for LLL streaming over the Internet, and (*iii*) different deployment scenarios of LLL streaming over the AWS Elemental Media Services. Please note that these media players are continuously being developed and improved. Thus, our experimental results reflect only a snapshot in time. The main point is to illustrate that media players are often designed and tuned differently to achieve different, sometimes contradictory, performance trade-offs.

TABLE VIII: Default parameters used in the experiments.

| Player | Parameters | Default Value |
|---|---|---|
| hls.js (LL-HLS) | maxBufferLength | 30 s |
| | maxBufferSize | 60 MB |
| | maxStarvationDelay | 4 s |
| | liveSyncDurationCount | 3 segments |
| | playlistReloadInterval | 1 s |
| dash.js (LL-DASH) | setFragmentLoaderRetryInterval | 1 s |
| | setFragmentLoaderRetryAttempts | 3 |
| | setBandwidthSafetyFactor | 0.9 |
| | setStableBufferTime | 12 s |
| | setBufferToKeep | 20 s |
| | setBufferTimeAtTopQuality | 30 s |
| | setLowLatencyEnabled | True |
| | setLiveDelayFragmentCount | 4 |
| Shaka (LL-DASH/LL-HLS) | bufferingGoal | 10 s |
| | rebufferingGoal | 2 s |
| | maxAttempts | 2 |
| | baseDelay | 1 s |
| | timeout | 0 |
| | backoffFactor | 2 |
| ExoPlayer (LL-DASH/LL-HLS) | Min_Loadable_Retry_Count | 3 |
| | Track_Blacklist_MS | 60 s |
| | Max_Buffer_MS | 3 s |
| | Buffer_For_Playback_MS | 2.5 s |
| | Buffer_For_Playback_After_Rebuffer_MS | 5 s |

Fig. 21 shows four deployment scenarios for media distribution using the AWS Elemental Media Services. These scenarios are summarized in Table IX and aim to investigate the combination of different AWS Elemental Media Services, *i.e.*, MediaLive, MediaPackage, MediaStore, Live and Delta, and AWS CloudFront CDN. Specifically, one workflow is fully deployed on-premises (Scenario 1) and three hybrid workflows (Scenarios 2–4) with encoder and packager or origin are deployed on-premises. These scenarios are similar to how Twitch uses AWS services to deliver content to its massive audience. For video contribution, a live feed at 1080p was created and transported from the live source to the encoder using RTMP.

TABLE IX: Testing open-source media players in different LLL streaming scenarios (*: on-premises). All scenarios use CloudFront CDN for delivery from the origin to the media player.

| Scenario | Encoder | Packager | Origin |
|---|---|---|---|
| 1 | Live* | Delta* | Delta* |
| 2 | Live* | MediaPackage | MediaPackage |
| 3 | Live* | MediaLive | MediaStore |
| 4 | Live* | MediaStore* | MediaStore |

These experiments can provide a general overview of how to decide and adjust an LLL streaming distribution workflow, including player parameters with the AWS Elemental Media Services, to meet the requirements of the target application. As the live source, we used the Big Buck Bunny (https://peach.blender.org/download/) sequence with a segment duration of five seconds and a chunk (or part) duration of one second (equal to 30 frames at 30 fps). Consistent with existing CMAF-based live services, we encoded the video with FFmpeg (H.264/AVC) into four video representations {360p@0.75 Mbps, 480p@1 Mbps, 720p@2 Mbps and 1080p@5 Mbps} [14] and the audio into one audio representation at 128 Kbps. The session duration was fixed at 600 seconds. We used the Google Chrome browser (v86) with Fetch API support to run the browser-supported media players (*i.e.*, dash.js, Shaka and hls.js), and Android (v10) to run ExoPlayer over WiFi home network conditions. The encoder, packager and origin server were in the USA, while the CDN edge and players were in Singapore. We fixed the target latency to five seconds, set the ABR scheme to heuristic hybrid-based (buffer + throughput), and we used LL-DASH or LL-HLS for the media distribution workflow, depending on the support in the player. For each experiment, we ran two tests—one with the default parameters (see Table VIII) and another with the well-tuned configurations of these parameters for LLL streaming—and re-visited the bandwidth measurement module [133] to avoid inaccuracies in measurements caused by CTE, as explained in Section VII-A.

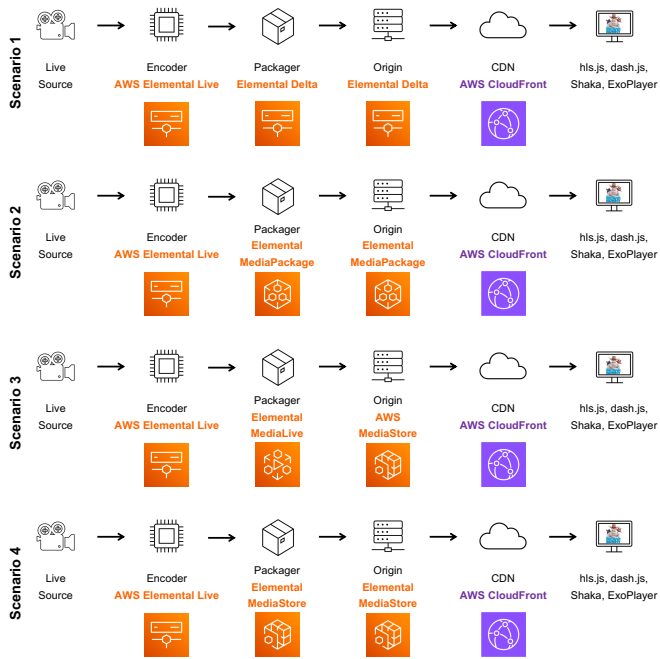

Fig. 21: Twitch-like LLL streaming distribution workflows using AWS Elemental Media Services and open-source media players.

Table X shows the results achieved by each media player in the four scenarios in terms of viewer QoE metrics [133]: E2E latency, selected quality (bitrate), rebuffering, startup delay and playback speed. The key takeaways of these experiments are the following. (Note that these are results at the time of the experiments. Newer or older versions of the media players may deviate.)

- Each player performed better with the well-tuned parameters than the default ones. Across all scenarios and player implementations, players with the well-tuned parameters achieved an average E2E latency of 5.11 seconds (near the target latency of five seconds), high quality, less rebuffering and low startup delay, compared to an E2E latency of 7.27 seconds for players with the default parameters. We note that players with the well-tuned parameters streamed at normal speed more than 85% of the time, with the remaining 15% either slower or faster during the whole live session. We expect this to have minimal impact on the visual quality.
- The players (of all implementations) achieved better results in Scenario 4, followed by Scenario 1, compared to other scenarios. This is because of the deployment of components on-premises compared to the cloud. Moreover, AWS Elemental Live with AWS Elemental MediaStore for distribution shows a good fit for LLL streaming. In Scenarios 2 and 3, AWS Elemental MediaPackage and MediaLive add more delay to the latency because of repackaging and transcoding.
- In all scenarios, dash.js achieves the best performance compared to the other open-source players, likely because dash.js is better tuned for LLL scenarios. We note that the other players are still working on improving their algorithms for LLL streaming.
- For media distribution at last-mile delivery, a player's network connection can significantly impact the latency. For example, a viewer may connect via a wired network at his/her home, via a WiFi access point, or using his/her mobile connection to access the content. Also, the proximity of the closest CDN server may impact the latency.

Overall, improvements in results can be attributed to the well-tuned player parameters, accurate bandwidth measurements, playback speed adjustments, and the location of delivery workflow components. We also tested with a chunk duration of 33 milliseconds and some adjustments at Cloud-Front, and we achieved better performance (not shown), which confirms that reducing latency requires an E2E workflow that is all streamlined and tuned together. We note that the media players are continuously being updated and their performance may vary across versions. Interested readers may also refer to [232] for an alternative study comparing the low-latency performance of several players (dash.js, hls.js, AVPlayer and Shaka player).

Finally, an E2E LLL streaming workflow can be further improved through analytics. One option is to use a centralized entity to collect data from each component in the delivery workflow by leveraging a Server and Network Assisted DASH (SAND) [34]–[36], [44], [105], [122], [168], [200]–[203] architecture or the Common Media Client Data (CMCD; CTA-5004) [37], [59] framework. Then, AI-based techniques can be used to make suitable decisions for better LLL delivery. For example, [37] demonstrated how the playback buffer status could be conveyed from the client to the server via CMCD to allow the server to dynamically allocate its outbound bandwidth according to each client's buffer level (*e.g.*, clients with critically low buffer levels are allocated a larger portion of the bandwidth). Similarly, the authors of [157] used three CMCD parameters, *i.e.*, top bitrate, device type and screen resolution, to construct a suitable bitrate ladder for each

IEEE COMMUNICATIONS SURVEYS AND TUTORIALS 29TABLE X: Results for different open-source media players in four scenarios. These tests were done using AWS Elemental Media Services with several open-source players (hls.js, dash.js, Shaka and ExoPlayer). The encoder, packager and origin server were in the USA, while the CDN edge and players were in Singapore. A fixed five-second target latency was used and the measurements were taken over 600 seconds. This highlights that different open-source media players behave differently in different LLL architectures. The result supports the idea that existing open-source media players need significant adjustments to adapt to LLL. It shows that the architecture choice, parameter selection and location of the streaming resources significantly affect a player's behavior.

| | Default Parameters | | | | | Well-Tuned Parameters | | | | |
|---|---|---|---|---|---|---|---|---|---|---|
| Player | E2E Latency (s) | Bitrate (Mbps) | Rebuffering Duration (s) | Startup Delay (s) | Playback Speed | E2E Latency (s) | Bitrate (Mbps) | Rebuffering Duration (s) | Startup Delay (s) | Playback Speed |
| | | | | | Scenario 1 | | | | | |
| hls.js | 8.54 | 3.50 | 37 | 3.34 | 1x | 5.11 | 3.70 | 21 | 2.10 | 0.5-1.5x |
| dash.js | 7.12 | 3.80 | 30 | 2.22 | 1x | 4.90 | 3.95 | 17 | 1.90 | 0.5-1.5x |
| Shaka | 10.20 | 4.00 | 45 | 4.14 | 1x | 5.20 | 3.75 | 19 | 2.40 | 0.5-1.5x |
| ExoPlayer | 11.40 | 2.90 | 24 | 4.09 | 1x | 5.80 | 3.20 | 14 | 2.60 | 0.5-1.5x |
| | | | | | Scenario 2 | | | | | |
| hls.js | 9.63 | 3.20 | 42 | 3.67 | 1x | 6.17 | 3.50 | 30 | 3.10 | 0.5-1.5x |
| dash.js | 7.90 | 3.50 | 34 | 3.50 | 1x | 5.60 | 4.00 | 22 | 2.50 | 0.5-1.5x |
| Shaka | 11.22 | 3.30 | 54 | 4.94 | 1x | 6.00 | 4.11 | 25 | 2.34 | 0.5-1.5x |
| ExoPlayer | 12.00 | 2.50 | 31 | 5.00 | 1x | 6.90 | 3.60 | 21 | 3.30 | 0.5-1.5x |
| | | | | | Scenario 3 | | | | | |
| hls.js | 8.11 | 3.60 | 32 | 3.00 | 1x | 5.17 | 4.10 | 24 | 2.10 | 0.5-1.5x |
| dash.js | 7.00 | 3.70 | 27 | 3.10 | 1x | 5.10 | 4.40 | 14 | 1.70 | 0.5-1.5x |
| Shaka | 9.50 | 3.60 | 39 | 3.30 | 1x | 5.70 | 4.20 | 16 | 2.80 | 0.5-1.5x |
| ExoPlayer | 10.00 | 3.00 | 18 | 3.83 | 1x | 6.50 | 3.90 | 12 | 2.00 | 0.5-1.5x |
| | | | | | Scenario 4 | | | | | |
| hls.js | 7.55 | 3.66 | 27 | 3.40 | 1x | 5.00 | 4.00 | 20 | 1.90 | 0.5-1.5x |
| dash.js | 6.90 | 3.98 | 27 | 3.00 | 1x | 4.90 | 4.30 | 13 | 1.50 | 0.5-1.5x |
| Shaka | 9.00 | 3.89 | 40 | 3.50 | 1x | 5.20 | 4.40 | 14 | 1.90 | 0.5-1.5x |
| ExoPlayer | 9.44 | 3.30 | 23 | 4.20 | 1x | 5.20 | 4.10 | 9 | 2.00 | 0.5-1.5x |

device type (*e.g.*, phone, TV or HDTV) at the server side. The companion framework, Common Media Server Data (CMSD; CTA-5006) [29], [30], [38], [39], [61], [134], [169], is designed to allow data to be conveyed from the server to the client to enhance media delivery. This could also open up new possibilities and use cases, such as multi-CDN performance monitoring and switching at the client [159].

## VIII. TWITCH'S GRAND CHALLENGE ON LOW-LATENCY LIVE STREAMING

In this section, we now present a practical case study based on the 2020 grand challenge organized by Twitch [206] to demonstrate the effectiveness of LL-DASH. We use this case study to highlight some of the best practices in prototyping and testing such an E2E solution. We also discuss key evaluation results and findings from the solutions submitted to this challenge to understand better how their respective implementations handle the challenging requirements of LLL streaming.

**About the Challenge.** In early 2020, Twitch partnered with the ACM Multimedia Systems Conference (MMSys) [206] and called for developers and researchers to design novel ABR schemes and player enhancements tailored toward LLL streaming. The challenge specifically targeted player enhancements as the player often has the biggest impact on latency in the LLL delivery workflow. We discuss further details of the challenge description and requirements in later sections.

**About Twitch and Motivation for the Challenge.** Twitch delivers interactive experiences to millions of users through LLL streaming. LLL streaming had an enormous impact on the platform design as users can get more engaged with streamers with real-time chat and other similar features and the development of new streaming experiences around this newfound interactivity. However, Twitch's feedback shows that the user experience may worsen significantly when streaming at low latencies. The root cause of this issue usually lies in the streaming client, particularly its ABR scheme, bandwidth measurement and playback speed control components, which have yet to evolve to meet the needs of various LLL streaming scenarios.

This issue is not exclusive to Twitch. The lack of suitable player enhancements (especially in the ABR scheme) that can balance various QoE metrics well amidst a low-latency live streaming environment is one of the main factors preventing LLL streaming from reaching widespread adoption and from reducing the E2E latency to even lower levels than what the industry has achieved so far.

### A. Detailed Requirements

The grand challenge listed the following requirements based on Twitch's experience with delivering LLL streams to users:
- Small Playback Buffer Size: The media player should maintain a playback buffer slightly smaller than the target latency. When the target latency is one second, the player will have less than one second to adapt to network conditions to avoid rebuffering or deviations from the target latency.
- Playback Speed Adjustments: The media player's playback speed controller should adjust the playback speed appropriately to meet the target latency constraint while not risking rebuffering events.
- Bandwidth Measurements: The media player should measure the available bandwidth accurately under varying

network conditions. Furthermore, it should make a timely request for the segments to be delivered via CTE as soon as possible. In other words, the media player should send the request for the segment as early as possible as long as it does not arrive before the first chunk is available. This effectively removes the forward-trip time from the segment request and allows a more stable download rate once the subsequent chunks are available for streaming.

- Robustness to Diverse Network Conditions: The media player should perform well across various network conditions. For example, the media player has to balance the number of bitrate switches against minimizing the number of rebuffering events while streaming at an acceptable quality and latency.
- Fairness among Media Players: Multiple media players sharing the same bottleneck network may compete for network resources. For proper operation, media players should respect the fairness constraint (refer to Section VII-B).

In summary, contestants entering the challenge needed to design an ABR scheme tailored toward LLL streaming with CMAF and CTE in the one-to-two-second latency range. The ABR scheme was expected to minimize the number of rebuffering events while maximizing bandwidth utilization and respecting the other challenge requirements mentioned above.

### B. Assets and Test Environment

In order to standardize and streamline development, Twitch provided a local, end-to-end testbed [205] to evaluate different implementations. The evaluation testbed is depicted in Fig. 22 and consists of three key components: the streaming engine, the network simulator and the metrics recorder, which are further detailed below. If the readers are interested in deploying the testbed and evaluating their customized ABR schemes, they may refer to the guidelines presented in [205].

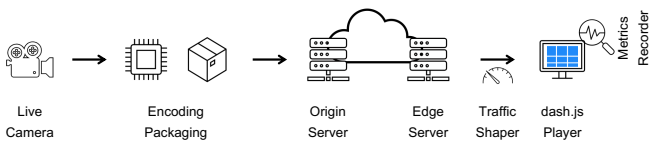

Fig. 22: Controlled experimental setup for the Twitch grand challenge [206].

- Streaming Engine: This delivers and plays a live streaming session frame-by-frame using CTE. It is comprised of four modules as follows: (*i*) A live source that captures the 720p rendition of the Big Buck Bunny (BBB) sequence[5]. (*ii*) An FFmpeg encoder with H.264/AVC codec that encodes the BBB sequence into three representations {360p@200 Kbps, 480p@600 Kbps, 720p@1,000 Kbps}, with a segment duration of 0.5 seconds and a chunk duration of 33 ms (one frame duration at 30 fps). The encoded streams are then packaged in CMAF format. (*iii*) An origin and edge DASH server, which uses CTE to deliver chunks to the player. (*iv*) A dash.js reference player (v3.0.1) [66] that streams the live content.
- Network Simulator: This throttles the bandwidth between the edge DASH server and the dash.js player according to predefined network profiles to mimic real-world scenarios. It is comprised of the following modules: (*i*) A Chrome dash.js player controlled by Puppeteer[6]. (*ii*) A node.js[7] application to run the tests with the respective network profiles. (*iii*) The Chrome DevTools protocol to throttle the bandwidth. (*iv*) Five different network profiles, namely {Cascade, Intra-Cascade, Spike, Slow-Jitters and Fast-Jitters} as shown in Table XI. The evaluation testbed waits for the player to play the highest-bitrate representation for ten seconds before the simulation begins (for warm-up stabilization purposes).
- Metrics Recorder: This listens to the dash.js player events and records metrics at each segment download, including selected bitrate, number of rebuffering events and their durations, number of bitrate switches, startup delay, E2E latency, bandwidth measurement and buffer level.

TABLE XI: Network profiles.

| Profile | Duration (s) | Avg. Bitrate (Kbps) | Switches (#) |
|---|---|---|---|
| Cascade | 150 | 680 | 5 |
| Intra-Cascade | 135 | 555 | 4 |
| Spike | 30 | 600 | 2 |
| Slow-Jitters | 30 | 600 | 6 |
| Fast-Jitters | 11.6 | 958 | 6 |

### C. Evaluation Criteria

All competing submissions were evaluated on a workstation running Ubuntu 19.10 with two virtual machines (VMs). The first VM ran the dash.js player executing in Chrome v81. The second VM hosted the FFmpeg encoder with CMAF packaging and the Python-based origin and edge DASH servers. For network emulation, the Chrome DevTools was used to throttle the bandwidth between the edge server and player according to the provided network profiles. The target latency was fixed to one second.

The submissions were evaluated based on the five network profiles (results were averaged over five runs for all profiles) and the following criteria: average selected bitrate, average E2E latency, average number of bitrate switches, average playback speed and average rebuffering duration. These measurements were fed into the QoE model given in (2) [133] to produce a single score, which was used to rank the submissions. The QoE model is defined as follows:

$$QoE_K = \sum_{i=1}^{K} \left( \alpha R_i - \beta E_i - \gamma L_i - \sigma |1 - P_i| \right) - \sum_{i=1}^{K-1} \mu H_i. \quad (2)$$

The list of the notations for the QoE model is given in Table XII. The weight factor of each QoE criterion was fixed based on subjective tests [229], [230] as follows: $\alpha$ = segment duration (0.5 seconds), $\beta$ = maximum encoding bitrate (1,000 Kbps), $\sigma$ = minimum encoding bitrate (200 Kbps), $\mu$ = 1 second, and $\gamma$ = {if $L \leq 1.1$ seconds then $\gamma = 0.05 \times$ minimum encoding bitrate, otherwise $\gamma = 0.1 \times$ maximum encoding bitrate}. The playback speed was usually: (*i*) *normal*: 1×, (*ii*) *fast*: *e.g.*, 1.5×, or (*iii*) *slow*: *e.g.*, 0.95×. Thus, if the playback speed were normal, then the playback speed penalty would be zero. Otherwise, the penalty was set to the minimum encoding bitrate level.

[5][Online] Available: https://peach.blender.org/download
[6][Online] Available: https://pptr.dev
[7][Online] Available: https://nodejs.org

TABLE XII: Notations of the QoE model.

| Notation | Meaning |
|---|---|
| $R$ | Bitrate selected (Kbps) |
| $E$ | Rebuffering duration (seconds) |
| $L$ | Live latency (seconds) |
| $H$ | Number of bitrate switches (#) |
| $P$ | Playback speed (*e.g.*, 1x, 0.95x, 1.05x) |
| $K$ | Total number of segments |
| $\alpha$ | Selected bitrate reward factor |
| $\beta$ | Rebuffering duration penalty factor |
| $\gamma$ | Live latency penalty factor |
| $\sigma$ | Playback speed penalty factor |
| $\mu$ | Bitrate switch penalty factor |

*D. Submissions, Results and Analysis*

In total, three solutions were submitted to the challenge: LoL (Low-on-Latency) [133], L2A (Learn-to-Adapt) [117], and Stallion (STAndard Low-Latency vIdeo cONtrol) [94]. More details on these solutions (including their source codes) can be found in their original papers. For LoL, we present the results of the learning-based ABR scheme. As the baseline, the default ABR scheme of dash.js (referred to as Dynamic, which uses a hybrid of BOLA [192] and throughput-based ABR) was used.

The average results achieved over five runs and all network profiles are highlighted in Table XIII. Tables XIV, XV, and XVI show the comparison results (average, median and standard deviation) for LoL vs. dash.js, L2A vs. dash.js and Stallion vs. dash.js, respectively, over five runs and all network profiles. Note that Stallion used a different target latency configuration (1.5 seconds instead of one second as used by all other solutions), contributing to its better rebuffering performance.

From the results, one key observation is that each solution has its own merits, such that it can achieve better results on specific QoE criteria or perform better under certain network conditions. For interested readers, the full details of the results can be found in [132].

Discussing more results, LoL achieved the highest average selected bitrate in all network profiles among the submissions due to its accurate bandwidth measurement (with an average accuracy of more than 96% across all network profiles). Moreover, LoL uses a learning-based ABR, which can track four dimensions with four different weights, including latency, bandwidth, buffer level and QoE, making the solution more sensitive to changes in any of these QoE criteria. Hence, it can quickly detect bandwidth changes and handle sudden bandwidth drops better. In contrast, dash.js, L2A and Stallion suffered from bandwidth overestimation or underestimation issues, resulting in lower bitrate selections, where inter-chunk idle times are more likely to occur. This follows our earlier discussion on how CTE may cause issues in existing bandwidth measurement techniques (refer to Fig. 20).

Twitch also observed that the performance of each solution varied under different network conditions. This suggested that different ABR solutions had their own strengths, and thus, using multiple ABR schemes, combined with an appropriate ABR *switching* module that selects the best ABR scheme based on changing requirements and environmental conditions, may help us leverage these solutions better. One can use machine learning techniques for such an ABR switching module.

In addition to objective tests, Twitch also conducted subjective tests internally to evaluate the submissions. The subjective tests aimed to answer the following questions: Do you think this solution could perform well on the diverse Twitch network? What is the potential for each design? Have the authors realized this potential? Based on the responses, Twitch concluded the following:

**For LoL**:
- *The bandwidth measurement appears to be accurate, where (i) the quick upswitches we observed are especially promising, and (ii) the authors refined ACTE [41] in response to pre-requests reducing burstability.*
- *The author's learning-based rule is innovative, but the hard-coded weights may be a fatal flaw. We believe it would be difficult or impossible to choose values that work for Twitch's diverse viewership statically.*

**For L2A**:
- *L2A is innovative, where (i) the authors framed adaptive streaming as a convex optimization problem, (ii) convex optimization seems well-suited to streaming, (iii) convex optimization is a fast expanding field, and L2A stands to benefit from this research.*
- *L2A has two main flaws: (i) the authors did not address playback rate increases causing rebuffering events, (ii) the authors used the default bandwidth estimator, which is not as accurate. However, those flaws are not insurmountable.*

**For Stallion**:
- *Stallion does make a favorable trade-off between bitrate and rebuffering duration, where the percentages seem large, but the actual values are small (e.g., about four seconds of rebuffering in the Cascade profile).*
- *Stallion's ideas are simple and effective: (i) we appreciated the attention on preventing playback rate increases from causing rebuffering, and (ii) it significantly reduces the number of switches. However, Stallion does not make significant improvements to the default dash.js ABR scheme (Dynamic). Hence, the root problem with the bandwidth measurement still remains.*

According to Twitch, and considering the output of both objective and subjective evaluations, picking the winner was not easy as the three solutions were similar in overall performance. In the end, Twitch announced the winners as follows: (1) L2A, (2) LoL and (3) Stallion. We note that the authors of LoL have considered Twitch's feedback and extended their solution in the follow-on LoL[+] [32]. LoL[+] avoids the hard-coded weights of its learning-based rule by implementing a heuristic-based algorithm to adjust the weights dynamically. It formulates the weight selection as an optimization weight assignment problem and solves it using a heuristic-based algorithm running in two states. (*i*) At the beginning of a live session (buffering-filling state), the weight selection algorithm uses a k-means++ mechanism [22] to find suitable initial weights for the considered features, *i.e.*, individual QoE metrics in (2) and bandwidth measurement. (*ii*) After the buffering-filling state (steady state), it uses a weight assignment algorithm that dynamically adjusts these weights at every segment download, allowing better performance. LoL[+] also includes a learning-based bitrate selection rule decoupled from the QoE function and selects based on individual QoE criteria in (2) instead. Finally, LoL[+] adopts a hybrid proactive approach to control the playback speed, jointly considering the current latency and current buffer level. Therefore, it aims to eliminate issues related to latency and buffer occupancy (*e.g.*, latency increase, rebuffering events and quality drops) before they occur. The detailed results and discussion of LoL[+] against LoL, L2A and Stallion are available in [32].

TABLE XIII: Average results over five runs and all network profiles. OPT-theoretical is the optimal solution with full knowledge of the system. These results come from a controlled experiment in the Twitch grand challenge using a simulator, which emulates network bandwidth variations. The performance of these proposed LLL ABR solutions and the baseline algorithm of dash.js was tested under five different network profiles.

| Solution | Avg. Selected Bitrate (Kbps) | Avg. E2E Latency (Second) | Avg. Rebuffering Duration (Second) | Avg. Buffer Level (Second) | Avg. Playback Speed (#) | Avg. Bitrate Switches (#) |
|---|---|---|---|---|---|---|
| dash.js | 276 | 1.17 | 11.02 | 0.43 | 1.08 | 4.12 |
| LoL | 697 | 1.87 | 20.86 | 0.49 | 1.21 | 5.16 |
| L2A | 440 | 1.69 | 16.89 | 0.42 | 1.12 | 6.56 |
| Stallion | 452 | 1.72 | 3.17 | 0.64 | 1.02 | 3.68 |
| OPT-theoretical (Oracle) | 679 | 1 | 0 | 1 | 1 | 4.60 |

TABLE XIV: Comparison of LoL and dash.js in terms of average results over five runs for all network profiles. Up/down arrows refer to an improvement/degradation in the metric [132].

| LoL vs. dash.js | Avg. Selected Bitrate (Kbps) | Avg. E2E Latency (Second) | Avg. Rebuffering Duration (Second) | Avg. Playback Speed (#) | Avg. Bitrate Switches (#) |
|---|---|---|---|---|---|
| Average | 152.3% ↑ | 59.5% ↓ | 89.3% ↓ | 12.4% ↓ | 25.2% ↓ |
| Median | 139.9% ↑ | 46.0% ↓ | 89.9% ↓ | 12.2% ↓ | 36.8% ↓ |
| Standard Deviation | -13.6% | 1165.0% | 46.3% | 80.1% | -41.8% |

TABLE XV: Comparison of L2A and dash.js in terms of average results over five runs for all network profiles. Up/down arrows refer to an improvement/degradation in the metric [132].

| L2A vs. dash.js | Avg. Selected Bitrate (Kbps) | Avg. E2E Latency (Second) | Avg. Rebuffering Duration (Second) | Avg. Playback Speed (#) | Avg. Bitrate Switches (#) |
|---|---|---|---|---|---|
| Average | 59.4% ↑ | 44.4% ↓ | 53.2% ↓ | 3.7% ↓ | 59.2% ↓ |
| Median | 53.7% ↑ | 45.8% ↓ | 55.1% ↓ | 4.3% ↓ | 73.7% ↓ |
| Standard Deviation | 211.1% | 235.8% | 72.8% | 43.2% | -32.8% |

TABLE XVI: Comparison of Stallion and dash.js in terms of average results over five runs for all network profiles. Up/down arrows refer to an improvement/degradation in the metric [132].

| Stallion vs. dash.js | Avg. Selected Bitrate (Kbps) | Avg. E2E Latency (Second) | Avg. Rebuffering Duration (Second) | Avg. Playback Speed (#) | Avg. Bitrate Switches (#) |
|---|---|---|---|---|---|
| Average | 63.8% ↑ | 47.0% ↓ | -71.3% ↑ | -6.0% ↑ | -10.7% ↑ |
| Median | 59.1% ↑ | 33.6% ↓ | -77.0% ↑ | -6.3% ↑ | -5.3% ↑ |
| Standard Deviation | 181.2% | 656.9% | 65.9% | 58.3% | -52.9% |

It is also worth noting that L2A and LoL[+] have been integrated into the dash.js reference player (from v3.2.0). Several recent works around low-latency live streaming (*e.g.*, [31], [45], [53], [161]) have conducted various comparison studies with these algorithms.

## IX. LESSONS LEARNED

In this survey, we explored the evolution of LLL streaming, dissecting workflows, protocols and technological advancements. Below, we synthesize the key lessons learned from this survey.

### A. Understanding Latency

- Latency as a goal in live streaming is not a one-size-fits-all metric. It varies based on the application, ranging from high latency (45+ seconds) for one-way streaming to near-real-time latency (sub-second) for interactive applications like cloud gaming and remote control of drones.
- E2E latency is influenced by multiple factors, including media content preparation (encoding, packaging, ingest), media content delivery (CDN distribution, last-mile delivery) and media content consumption (buffering, decoding, rendering).
- Achieving low latency often involves trade-offs with other factors such as video quality, encoding efficiency and playback stability. For example, using shorter segments may reduce latency but decrease encoding efficiency and increase the risk of rebuffering.

### B. Key Enablers for LLL Streaming

- Breaking media into smaller chunks (*e.g.*, using CMAF) allows for faster delivery and reduces the dependency on full segment availability. This is crucial for achieving low latency. Chunking retains some of the efficiencies of longer segments while achieving the low latency of very short segments.
- Protocols like CTE in HTTP/1.1 and server push in H2 and H3 enable the delivery of media chunks as soon as they are available, further reducing latency.
- Extensions like LL-DASH and LL-HLS have been developed to support chunked delivery and reduce the latency in HAS systems.
- While HTTP-based protocols (LL-DASH, LL-HLS) remain popular for scalability, QUIC/H3 eliminates head-of-line blocking and enables sub-500 ms latency, critical for interactive applications.
- RTMP and SRT remain entrenched in contribution/ingest due to their reliability despite the incompatibility with modern ABR streaming.

### C. Protocols and Workflows

- Push-based protocols (*e.g.*, RTMP, SRT, RIST) are commonly used for media contribution and ingest, while pull-based protocols (*e.g.*, DASH and HLS) are used for distribution. Each has its advantages and limitations in terms of latency, scalability and reliability.
- P2P systems can help scale live streaming by distributing content among peers, but they introduce challenges in terms of complexity, latency and reliability, especially under dynamic network conditions. They also suffer from difficulties such as large-scale service quality monitoring.
- New protocols like WebRTC and MOQ are being explored for ultra-low-latency streaming, especially for interactive applications like cloud gaming and eSports.

### D. Media Player Enhancements

- Accurate bandwidth measurement is critical for ABR streaming. Traditional methods may fail in low-latency scenarios due to chunked delivery, leading to overestimation or underestimation of the available bandwidth. Advanced techniques like ACTE and LoL have been developed to address this issue.
- ABR algorithms need to be redesigned for LLL streaming, as traditional algorithms may not perform well with small buffer sizes and variable playback speeds. Learning-based approaches (*e.g.*, L2A, LoL and LoL$^+$) and heuristic-based approaches (*e.g.*, Stallion) have shown promise in balancing quality, latency and rebuffering.
- Adjusting media playback speed dynamically can help maintain low latency and prevent buffer underruns. However, this must be done carefully to avoid noticeable artifacts and maintain synchronization between various media tracks.
- In LLL streaming, buffer sizes are smaller, making buffer management more challenging. Strategies like sticking to the live edge and resilience to late/missing segments are crucial for maintaining a smooth playback.

### E. CDN and Network Optimizations

- Efficient caching policies and prefetching strategies are essential for reducing latency in CDN distribution. Techniques like request coalescing and partial cache sharing can help minimize cache misses and improve delivery speed.
- Sudden spikes in viewer numbers (*e.g.*, during live sports events) can strain CDNs. Solutions like origin shields and multi-CDN distribution can help handle flash crowds and maintain low latency.
- The last-mile network (*e.g.*, home WiFi, 5G) plays a significant role in latency. Advances in 5G/6G and edge computing are expected to further reduce last-mile latency and improve the quality of live streaming.

### F. QoE in LLL Streaming

- Key QoE metrics for LLL streaming include E2E latency, startup delay, quality, rebuffering events and playback speed. While rebuffering events are the most detrimental to QoE, all these metrics are interdependent, and optimizing one often impacts the others.
- Both subjective (user studies) and objective (mathematical models like ITU P.1203) methods are used to evaluate QoE. The Yin model [230] and other QoE models help quantify the viewer experience based on a combination of these metrics.
- Achieving high QoE in LLL streaming requires balancing multiple factors, such as encoding quality, bitrate adaptation and playback stability. For example, higher encoding bitrates improve quality but increase latency, while aggressive bitrate adaptation can lead to frequent rebuffering.

### G. Case Study: Twitch's Grand Challenge

- Twitch's grand challenge highlighted the need for small playback buffers, accurate bandwidth measurement, robustness to network variations and fairness among media players in LLL streaming.
- Solutions like LoL, L2A and Stallion demonstrated the importance of learning-based ABR schemes, playback speed control and dynamic buffer management in achieving low latency while maintaining high QoE.
- The challenge showed that no single ABR algorithm performs best in all scenarios. A combination of heuristic-based and learning-based approaches, along with adaptive playback speed control, is necessary to handle diverse network conditions and viewer requirements.

### H. Industry Best Practices

- Achieving low latency requires optimizing every component of the media delivery pipeline, from encoding and packaging to CDN distribution and playback buffering.

- A poorly configured component can significantly impact the overall latency.
- The industry is moving toward interoperable solutions like CMAF and cross-format delivery to reduce the complexity of supporting multiple streaming protocols (*e.g.*, DASH and HLS) and improve cache efficiency.
- Ongoing standardization efforts, such as LL-DASH, LL-HLS and MOQ, are crucial for ensuring compatibility and scalability across different platforms and devices.

## X. Conclusion and Future Directions

Live media streaming over the Internet has exploded in popularity. A generation of "cord cutters," walking away from expensive TV subscriptions, combined with a sizeable growth in Internet speeds and new opportunities for interactivity within live broadcasts, have driven many content producers and broadcasters to online distribution. Media streaming platform heavyweights such as Twitch, YouTube, Meta, TikTok and X are making a significant push into LLL streaming services by encouraging their viewers to generate new live content and hosting major live events. That is to say, delivering LLL streams over the Internet is essential for many OTT applications and use cases such as live sports, eSports, live user-generated content, online gaming and just about anyone who wants to offer a great user experience at the lowest possible latency. As the drive toward interactivity increases, so does the demand to provide reduced latency, lower rebuffering, and high and stable quality, all of which are essential for a good QoE. To that effect, the entire end-to-end architecture and the components involved in the workflow must be well understood, which is one of the main motivations of this survey.

This survey provided a comprehensive overview of the components, protocols and technologies involved in LLL streaming. Through exploring key technologies such as chunked encoding/packaging, chunked delivery and the extensions for the existing HAS protocols, we gained insights into the advancements that enable efficient low-latency media delivery. Additionally, examining real-time communication protocols such as WebRTC and new proposals such as MOQ shed light on the potential for enhancing interactivity in LLL streaming. Furthermore, evaluating media player components, and bandwidth measurement and playback speed control algorithms contributed to our understanding of the challenges and opportunities in achieving low-latency playback. Finally, the lessons learned from Twitch's grand challenge on LLL streaming highlighted the importance of accurate bandwidth measurements, the potential of innovative approaches such as joint rate adaptation and playback speed control, and the need for adaptability to diverse viewers' requirements.

While we covered the fundamental technologies and standards in this survey, there is still significant room for improvement in the field of LLL streaming. Specifically, we expect future research and development to focus on the following:

- Leveraging analytics and AI-powered solutions in the context of LLL streaming involves leveraging the power of machine learning techniques to analyze user engagement, behavior and feedback data, enabling dynamic resource allocation, latency reduction, content-aware encoding and delivery, and real-time decision-making. By understanding user preferences and behavior patterns, AI-powered solutions can adapt the streaming workflow to deliver a personalized and high-quality viewing experience, while predictive analytics can preemptively optimize the delivery process based on anticipated user requirements, ultimately reducing latency. Additionally, real-time decision-making and dynamic resource allocation based on network conditions, server load and viewer demand can ensure a consistent QoE and minimize latency, paving the way for a new era of data-driven, adaptive and intelligent LLL streaming.
- Demystifying existing LLL streaming solutions involves disseminating knowledge and best practices to the broader community, facilitating a deeper understanding of the underlying technologies and methodologies. By sharing insights into successful LLL streaming implementations, including technical details, optimization strategies and performance benchmarks, the industry can collectively benefit from a more transparent and collaborative approach to addressing latency challenges. This knowledge-sharing initiative can encompass the documentation of case studies, performance evaluations and real-world deployment scenarios, enabling stakeholders to gain valuable insights into the practical considerations and trade-offs involved in LLL streaming.
- Further exploration of ABR schemes and buffer-management strategies explicitly tailored for low-latency scenarios includes investigating the impact of network conditions, user preferences and content characteristics on ABR decisions and developing dynamic buffer-management techniques to optimize playback latency and quality.
- The exploration of novel content delivery architectures, such as edge computing and distributed caching mechanisms, holds promise for minimizing latency and improving scalability in LLL streaming. Investigating the trade-offs and performance implications of these architectures under varying network conditions and viewer demands will be crucial for advancing the state-of-the-art in low-latency media delivery.
- Due to the need for high interactivity, latency is a critical metric in immersive applications (*e.g.*, VR/AR/MR applications) using 360-degree and volumetric media (*e.g.*, 3DoF and 6DoF videos) and studying LLL streaming solutions for such applications is vital.
- QUIC is becoming a great alternative for delivering media against TCP, the de facto layer running beneath most HTTP protocols since day one. Leveraging QUIC for live streaming could prove to be a game-changer for the field in terms of performance and stability.
- Leveraging the CMCD and CMSD standards may improve the user experience in LLL streaming. Media players and CDNs can use these specifications to exchange auxiliary information within object requests,

- which may be helpful for log analysis, quality monitoring and delivery enhancements.
- Developing QoE management frameworks tailored for LLL streaming will be essential for addressing the unique challenges associated with interactive and immersive media experiences. This includes investigating the impact of latency, rebuffering and quality fluctuations on user engagement and satisfaction, and developing metrics and models for evaluating QoE in LLL streaming systems.
- Developing real-time AIGC algorithms that minimize latency and dynamically adjust quality based on network conditions is essential. Further research is needed to enhance personalization and interactivity, enabling AI systems to create tailored and engaging content for viewers.
- Developing an energy-efficient streaming pipeline to reduce the power consumption of live video streaming over wireless, especially 5G and 6G networks.
- Future research in 5G and 6G for live video streaming should focus on developing ultra-low latency, high-bandwidth technologies and leveraging AI for dynamic network management and content adaptation. Additionally, enhancing mobility support, ensuring seamless handovers, and integrating with emerging technologies like AR/VR and IoT will be crucial for providing immersive, reliable, and high-quality streaming experiences.

These future research directions represent exciting opportunities for advancing the capabilities and performance of LLL streaming systems. By addressing the challenges in low-latency media contribution, ingest, delivery and playback, researchers can help unlock the full potential of LLL streaming in various application domains.

ACKNOWLEDGMENTS

This research has been supported by Singapore Ministry of Education Academic Research Fund Tier 2 under MOE's official grant number T2EP20221-0023.